\documentclass[final,5p,times,twocolumn]{elsarticle}

\usepackage{graphicx}

\usepackage{amsmath}

\usepackage{lineno}

\journal{Nuclear Instruments and Methods in Physics Research A}

\begin{document}

\begin{frontmatter}



\title{High density cluster jet target for storage ring experiments}


\author[adr1]{A. T\"aschner\corref{cor1}}
\ead{taschna@uni-muenster.de}
\author[adr1]{E. K\"ohler}
\author[adr1]{H.-W.~Ortjohann}
\author[adr1]{A. Khoukaz}

\cortext[cor1]{Corresponding author.}

\address[adr1]{Institut f\"ur Kernphysik, Westf\"alische
  Wilhelms-Universit\"at M\"unster, D-48149 M\"unster, Germany}

\begin{abstract}
The design and performance of a newly developed cluster jet target
installation for hadron physics experiments are presented which, for
the first time, is able to generate a hydrogen cluster jet beam with
a target thickness of above $10^{15}\,\mathrm{atoms/cm}^2$ at
a distance of two metres behind the cluster jet nozzle. The
properties of the cluster beam and of individual clusters themselves
are studied at this installation. Special emphasis is placed on
measurements of the target beam density as a function of the relevant
parameters as well as on the cluster beam profiles. By means of a
time-of-flight setup, measurements of the velocity of single clusters
and velocity distributions were possible. The complete installation,
which meets the requirements of future internal fixed target
experiments at storage rings, and the results of the systematic
studies on hydrogen cluster jets are presented and discussed.
\end{abstract}


\end{frontmatter}


\section{Introduction}

\begin{table*}
  \begin{center}
   \begin{tabular}{l|ccc|c}
                          & PROMICE/WASA    & E835                & ANKE and COSY-11      &             \\
                          & (CELSIUS)       & (FERMILAB)          & (COSY)                & M\"unster   \\
                          & \cite{Ekstroem1995}   & \cite{Allspach1998} & \cite{Dombrowski1997a} & (this work) \\
     \hline
     nozzle diameter      & $< 100\,\mu\mathrm{m}$ & $37\,\mu\mathrm{m}$ & $11-16\,\mu\mathrm{m}$ & $11-28\,\mu\mathrm{m}$ \\
     gas temperature      & $20-35\,\mathrm{K}$   & $15-40\,\mathrm{K}$ & $22-35\,\mathrm{K}$    & $19-35\,\mathrm{K}$\\
     gas pressure         & $1.4\,\mathrm{bar}$   & $< 8\,\mathrm{bar}$ & $18\,\mathrm{bar}$     & $> 18\,\mathrm{bar}$ \\
     distance from nozzle & $0.325\,\mathrm{m}$   & $0.26\,\mathrm{m}$  & $0.65\,\mathrm{m}$     & {\bf $2.1\,\mathrm{m}$} \\
     target thickness     & $1.3\times10^{14}\,\mathrm{cm}^{-2}$
                          & $> 2\times10^{14}\,\mathrm{cm}^{-2}$
                          & $\gg 10^{14}\,\mathrm{cm}^{-2}$
                          & {\bf $\ge 10^{15}\,\mathrm{cm}^{-2}$} \\
   \end{tabular}
  \end{center}
  \label{tab:targets}
  \caption{Comparison of the typical operation parameters and
  achieved target densities of different cluster jet targets. The
  densities observed with the M\"unster setup (right hand column) are obtained at a distance of
  more than two metres behind the nozzle.}
\end{table*}

Internal target beam facilities at storage rings play an important 
role in atomic, nuclear,
and particle physics. They allow for precision experiments
in combination with high luminosities and low physical background. 
In contrast to standard fixed-target installations, where the beam 
hits a target only once, in storage ring experiments  the accelerated 
beam particles  traverse the target material after each round-trip, 
typically with a repetition rate of 
$10^{5}-10^6\,\mathrm{s^{-1}}$. In order to make the best possible
use of the stored beam the stationary 
target must be rather thin, i.e. typically only 
$10^{12}-10^{15}\,\mathrm{nuclei/cm^2}$
thick. This is particularly important if beams of rare particles,
such as antiprotons, are used. In these cases the limited production rate
should not be exceeded by the consumption rate.
Further limitations of the interaction rate are introduced by the detector
setup and the data acquisition system.
Such a thin target can be 
realized by gas at rest, or beams of gas, clusters, or pellets.
Depending on the experimental situation, at each passage through 
the interaction point only a
fraction of the stored particles interacts with the target while 
the
other projectiles remain in the beam. 
However, some beam losses
will happen through scattering. Single scattering losses,
where beam particles are kicked out of the acceptance in
one collision, can be minimized by a large angular 
acceptance at the interaction point. By using beam
cooling devices, such as electron or stochastic cooling, 
multiple scattering losses accumulated over many turns
can be compensated.
Depending
on the conditions, the injected accelerator beam can be used for
cycle times from minutes to hours. 

The luminosity $L$ of an internal target experiment is given by the number of
circulating beam particles $N_\mathrm{C}$, the revolution frequency
$f$ of the stored particles, and the target thickness $n_\mathrm{t}$:
\begin{equation}
  L = n_\mathrm{t}\, N_\mathrm{C}\, f
\end{equation}
Typical values for these parameters in hadron physics experiments are
about $10^{10}-10^{11}$ circulating particles and an areal target thickness
on the order of $10^{12}-10^{15}\,\mathrm{atoms/cm^2}$~\cite{Lehrach2003}.

Although the requirements for the internal targets might differ from
experiment to experiment, some of them are more general:
The target thickness should be constant over time to
avoid luminosity fluctuations in the data acquisition system. 
Furthermore the target material should
show a homogeneous density distribution
at the interaction point. It should be of 
highest purity in order to avoid background reactions originating
from unwanted target isotopes.
The areal target thickness should be 
adjustable over a wide range in order to be able to match the
requirements of the physics programme and the specific detection
system.
In the absence of vertex tracking
detectors, the track and momentum reconstruction for ejectiles is
aided by a precise knowledge of the interaction region. This commonly
requires the geometrical target extension to be as small as possible
with a sharp boundary between the target and the surrounding vacuum.
The
above mentioned requirements might, of course, be very challenging if
a special element has to be used as a target material.

In case of experiments where only a low target thickness, on the
order of $10^{12}\,\mathrm{atoms/cm^2}$, is desired or needed
(supersonic) gas jets, produced by the expansion of gases through
fine nozzles into vacuum, are typically used. While, in principle,
all gaseous target materials could be used and the target thickness
would display no time structure, this type of target has the
disadvantage of generating a target beam with a large geometrical
divergence. In addition, the spatial target density distribution is
locally homogeneous, apart from the distinct structures caused by
shock fronts in the immediate vicinity of the nozzle. One major
disadvantage of this type of target is that the nozzle has to 
be placed close to the interaction point ($\ll 1\,\mathrm{m}$)
if high target densities 
($\ge 10^{12}\,\mathrm{atoms/cm^2}$) are required. This commonly 
introduces a high gas load to the storage ring vacuum.

On the other hand, if a high effective target thickness of up to
$10^{16}\,\mathrm{atoms/cm^2}$ is desired, a pellet
target~\cite{Ekstroem1996} can be used. In this kind of target
micrometer-size frozen spheres, with an individual target thickness
on the order of e.g.\ $10^{20}\,\mathrm{atoms/cm^2}$, are generated
from the target materials.
The use of such microspheres
introduces a very distinct time structure in the target beam. The
mean effective target thickness can be modified, either by varying
the distance of the pellets, or by changing the pellet size.
This leads to changes in the target
thickness of up to an order of magnitude. Details about this type of
target can be found in Ref.~\cite{nordhage}.

In contrast, cluster jet targets provide a freely adjustable target
thickness of up to about
$10^{15}\,\mathrm{atoms/cm^2}$~\cite{Taeschner2007} even at larger distances from the nozzle. 
They produce
small nanoparticles from cooled gases or liquids via expansion in
Laval-type nozzles, by either condensation of the gas or by breaking
up the liquid into a spray of tiny droplets. These clusters typically
consist of $10^3-10^6$ molecules~\cite{Knuth1995,General2008}. Different to
pellet targets, the random nature of the cluster production
introduces a homogeneous spatial distribution and no distinct time
structure. All gaseous target materials can be used for this kind of
target. The cross-section of the cluster beam at the interaction
point exhibits a well defined boundary so that the interaction region
is well defined. This property has the
important consequence that the beam can pass several metres through
an ultra high vacuum with constant angular divergence defined by the
orifices used for the cluster beam preparation.

Future internal fixed-target experiments at storage rings require
predominantly high density hydrogen or deuterium targets as effective
proton and deuteron/neutron beams. Due to their various advantages,
cluster jet targets are well established in high precision
experiments~\cite{Allspach1998,Dombrowski1997a,ANKE}. Typical
operation parameters of some of the targets used at the CELSIUS
ring~\cite{Ekstroem1995}, at the Fermilab Antiproton
Accumulator~\cite{Allspach1998}, and at
COSY~\cite{Dombrowski1997a,Stein} are listed in
Table~\ref{tab:targets}. Due to the finite angular divergence of the
cluster beam, the volume target density decreases quadratically with
the distance from the nozzle. At these experiments the distances
between the cluster production nozzles and the interaction points
range between 0.26~m and 0.65~m, resulting in hydrogen target
densities of several $10^{14}\,\mathrm{atoms/cm^2}$. Future
facilities for hadron physics experiments will be typically $4\pi$
detectors, where the interaction point between the stored beam and
the target beam is surrounded by onion-like layers of detectors
embedded in a magnetic field, allowing a momentum reconstruction of the
charged ejectiles. One prominent example of such an upcoming
installation is the planned PANDA detector at FAIR in
Darmstadt~\cite{PANDA2005}. The proposed detector concept requires
that the target source must be mounted outside the coils of the magnet and its
surrounding iron yoke, with the result that the distance from the
nozzle to the interaction point will increase from around
$0.5\,\mathrm{m}$, as in current targets, to more than 2 
$\mathrm{m}$. For cluster jet targets, this increase leads, for
geometrical reasons, to a decrease in the target thickness by
approximately one order of magnitude. Because of this effect, new
studies on higher cluster jet target densities were started in
M\"unster. We here present a newly developed cluster target device
which, for the first time, is able to provide hydrogen cluster beam
densities on the order of $10^{15}\,\mathrm{atoms/cm^2}$ at a distance of
more than two metres from the nozzle. This achievement opens the way
to use the broad advantages of cluster jet beams in future hadron
physics experiments with large $4\pi$ detectors.

\begin{figure}[b!]
\includegraphics[width=\columnwidth]{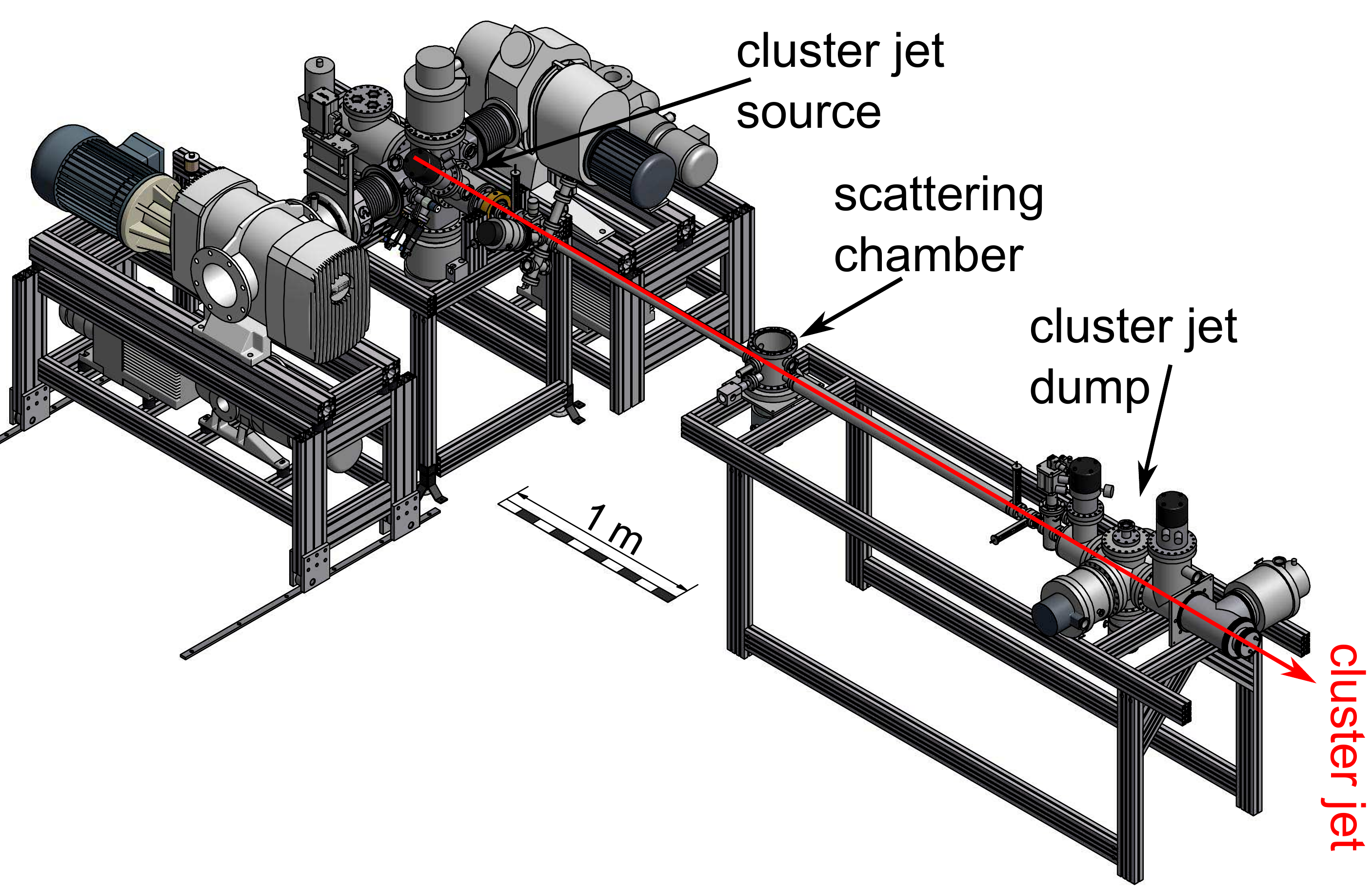}
\caption{View of the complete setup, with the cluster source and its
pumping system on the left, the scattering chamber with beam
diagnostics in the middle, and the cluster beam dump on the right.}
\label{fig:CompleteTarget}
\end{figure}

\section{Experimental setup}

A complete target station was designed and built up which allows one
to test improved cluster sources in a geometry which matches that of
an internal storage ring experiment with a $4\pi$ detector. Since the
dimensions of the future PANDA experiment are 
specified~\cite{PANDA2005}, the distances relevant to this target
station were used for the setup. In Fig.~\ref{fig:CompleteTarget} a
CAD view of the experimental setup is shown, with the cluster source
and its pumping system on the left, a scattering chamber with beam
diagnostics in the middle, and the cluster beam dump on the right.
The distance of $2.1\,\mathrm{m}$ between the nozzle inside the
cluster source and the centre of the scattering chamber is the same
as the one planned for the PANDA experiment. Furthermore, the
distance between the scattering chamber and the cluster beam dump
also matches the one from PANDA. In this way the data obtained can be
directly translated to the situation in the planned storage ring
experiment.

\begin{figure}[b!]
\includegraphics[width=0.8\columnwidth]{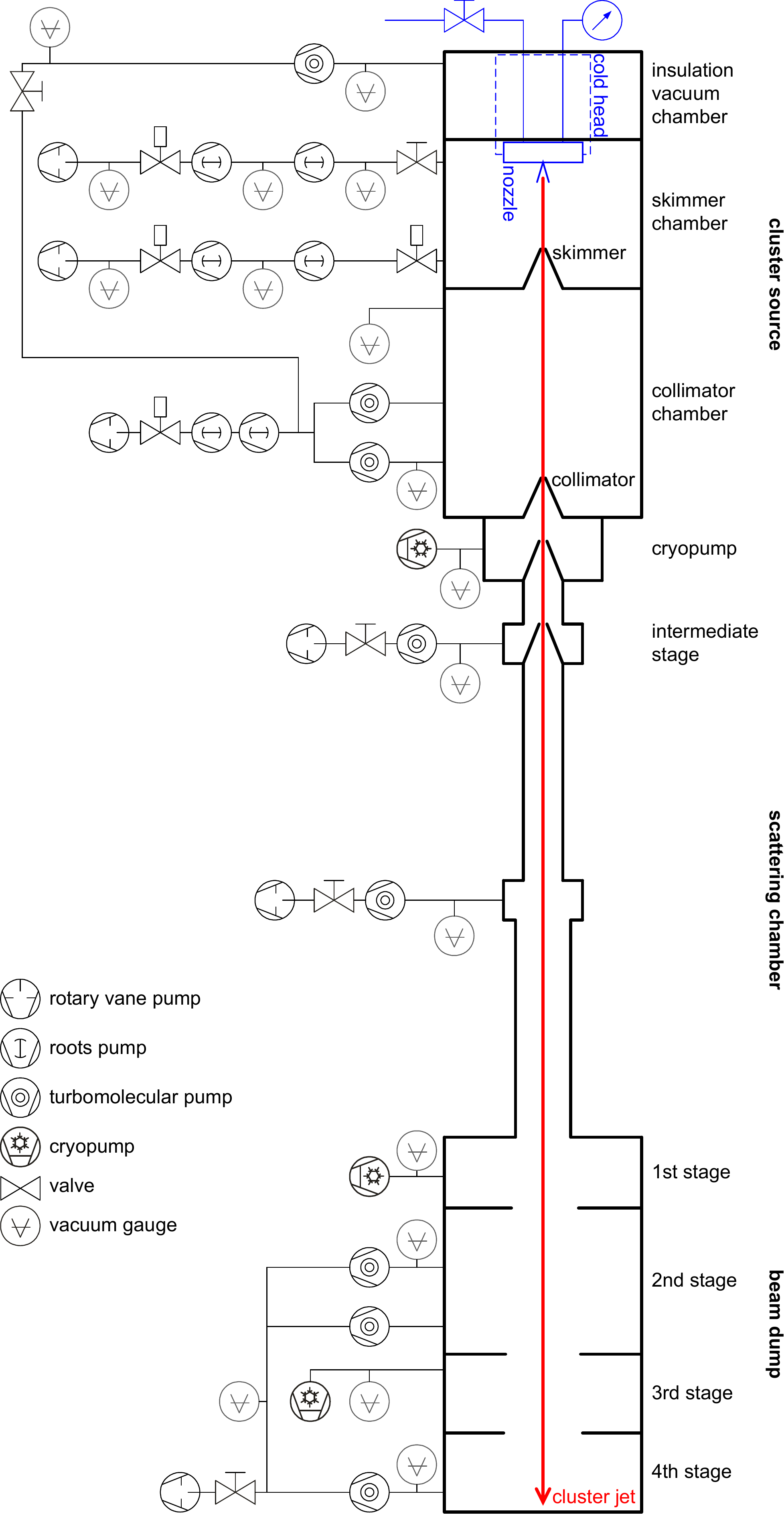}
\caption{The vacuum system of the M\"unster cluster jet target
setup.} \label{fig:VacuumSystem}
\end{figure}

In the source the clusters are generated and the shape of the cluster
jet is prepared. The cluster source consists of different vacuum
chambers forming a differential pumping system in order to reduce the
gas flow from the source into the interaction region. This is of
great importance in an internal storage ring experiment in order to
guarantee a low residual gas load on the accelerator beam line, as
well as to minimize reactions with the background gas. The scattering
chamber installed is equipped with a beam diagnostic system for the
determination of target thickness and geometry. The attached
cluster beam dump is designed to catch the cluster jet beam and to
minimize the back flow of gas to the scattering chamber. In
Fig.~\ref{fig:VacuumSystem} the complete design of the vacuum system
of the target setup is shown. The individual parts will be discussed
in the following sections. A list of the most important pumps, and
typical pressures observed during the experiments in the different
pumping stages, is compiled in Table~\ref{tab:Pumps}. For the
investigations presented in this work, the main emphasis was placed
on achieving highest target thicknesses and the measurement of the
cluster beam properties. Thus, with exception of the skimmer and the
collimator, the orifices connecting neighbouring vacuum chambers have
been chosen rather generously in diameter in order to avoid hitting
them when changing the cluster beam diameter through the use of
different collimator sizes. Therefore, the vacuum conditions
presented here are expected to be improved significantly when the
cluster beam size is fixed and appropriate orifice diameters are chosen.
\begin{table}[h!]
\small
\begin{tabular}{lll}
  Pumping stage    & Pumps                                       & Typical pressure  \\ \hline \hline
  Insulation       & Turbo pump: $370\,\mathrm{l/s}$             & $10^{-5}\,\mathrm{mbar}$\\
  vacuum chamber   & (Leybold Turbovac 361)                      & \\ \\
  Skimmer          & Roots pump: $3000\,\mathrm{m}^3\mathrm{/h}$ & $8\times10^{-2}\,\mathrm{mbar}$ \\
  chamber          & (Leybold Ruvac RA3001)                      & \\
                   & Roots pump: $2000\,\mathrm{m}^3\mathrm{/h}$ & \\
                   & (Leybold Ruvac WSL2001)                     & \\ \\
  Collimator       & Turbo pump: $2 \times 900\,\mathrm{l/s}$    & $2\times10^{-4}\,\mathrm{mbar}$\\
  chamber          & (Leybold Turbovac 1000 C)                   & \\ \\
  Cryopump stage   & M\"unster type cryopump                     & $2\times10^{-5}\,\mathrm{mbar}$\\ \\
  Intermediate     & Turbo pump: $110\,\mathrm{l/s}$             & $10^{-5}\,\mathrm{mbar}$\\
  stage            & (Leybold Turbovac 150)                      & \\ \hline \\
  Scattering       & Turbo pump: $340\,\mathrm{l/s}$             & $3\times10^{-5}\,\mathrm{mbar}$ \\
  chamber          & (Leybold Turbovac 360)                      & \\ \\ \hline
  1st beam dump    & M\"unster type cryopump                     & $6\times10^{-6}\,\mathrm{mbar}$\\ \\
  2nd beam dump    & Turbo pump: $900\,\mathrm{l/s}$             & $7\times10^{-6}\,\mathrm{mbar}$\\
                   & (Leybold Turbovac 1000 C)                   & \\
                   & Turbo pump: $370\,\mathrm{l/s}$             & \\
                   & (Leybold Turbovac 361 C)                    & \\ \\
  3rd beam dump    & M\"unster type cryopump                     & $3\times10^{-5}\,\mathrm{mbar}$\\ \\
  4th beam dump    & Turbo pump: $900\,\mathrm{l/s}$             & $8\times10^{-5}\,\mathrm{mbar}$\\
                   & (Leybold Turbovac 1000 C)                   &
\end{tabular}
\caption{Vacuum pumping system of the cluster jet target with
pressures for operation at highest target thickness and therefore
highest gas flow through the nozzle.} \label{tab:Pumps}
\end{table}

\begin{figure}
\center
\includegraphics[width=0.9\columnwidth]{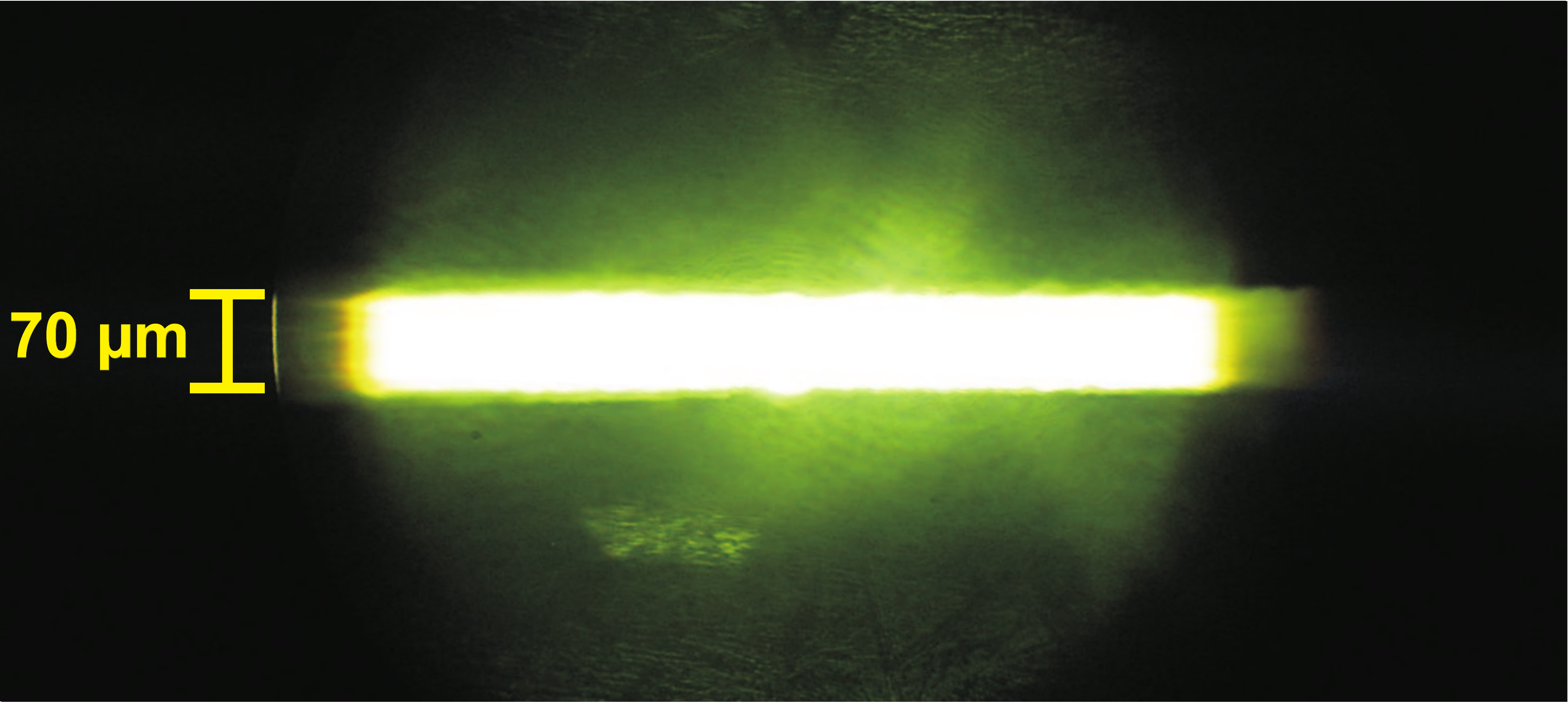}
\caption{Microscopic view of the orifice of a slit collimator
produced by a laser-cut method.} \label{fig:lasercollimator}
\end{figure}

More detailed investigations on the minimization of the residual gas
flow to the scattering chamber are currently under
evaluation~\cite{Taeschner2011}. Apart from the standard way to
suppress the gas background by reducing the orifice diameters between
the vacuum chambers, there is also the possibility to use cluster
beams with optimized cross-sections.
At a given target thickness the flow rate of the cluster beam can be minimized
by reducing the beam size transverse the direction of the ion or electron beam.
Assuming an ion beam size of one millimetre and a cluster beam size of
ten millimetres the gas flow can be reduced by almost one magnitude using a
cluster beam with a rectangular cross-section of 
$1\,\mathrm{mm}\times 10\,\mathrm{mm}$ instead of a circular one.
This concept has been proposed before, e.g. for the UA6 experiment at CERN
using a polarized atomic beam target \cite{DickKubischta1988}, and used
successfully with a cluster jet target at the COSY-11 experiment
\cite{Taeschner2006}.
First tests on
this idea have been carried out at the cluster target installation by
using specially shaped collimators produced by laser cutting. In
Fig.~\ref{fig:lasercollimator} a microscopic view of such an orifice
is shown with a size of $70\,\mathrm{\mu m}\times 860\,\mathrm{\mu
m}$, resulting in a cluster beam with a length of $\sim 13\,\mathrm{mm}$
in the ion beam direction and $\sim 1.5\,\mathrm{mm}$ perpendicular to it.
\begin{figure}[t!]
\center
\includegraphics[width=0.7\columnwidth]{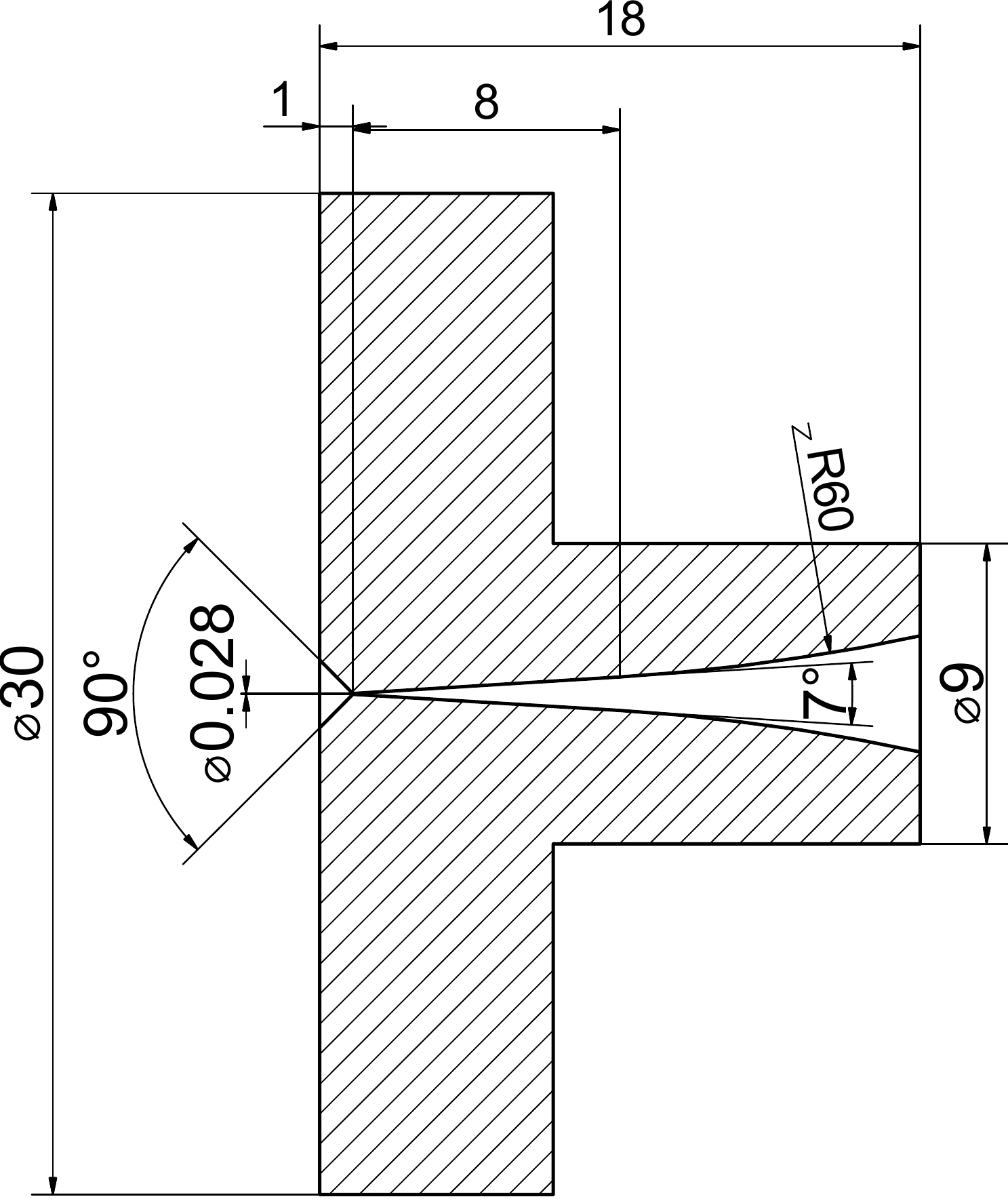}
\caption{Cross-section of the copper Laval nozzle used. This was
manufactured in the CERN workshop.} \label{fig:Nozzle}
\end{figure}
\subsection{Cluster source}

\begin{figure}
\center
\includegraphics[width=\columnwidth]{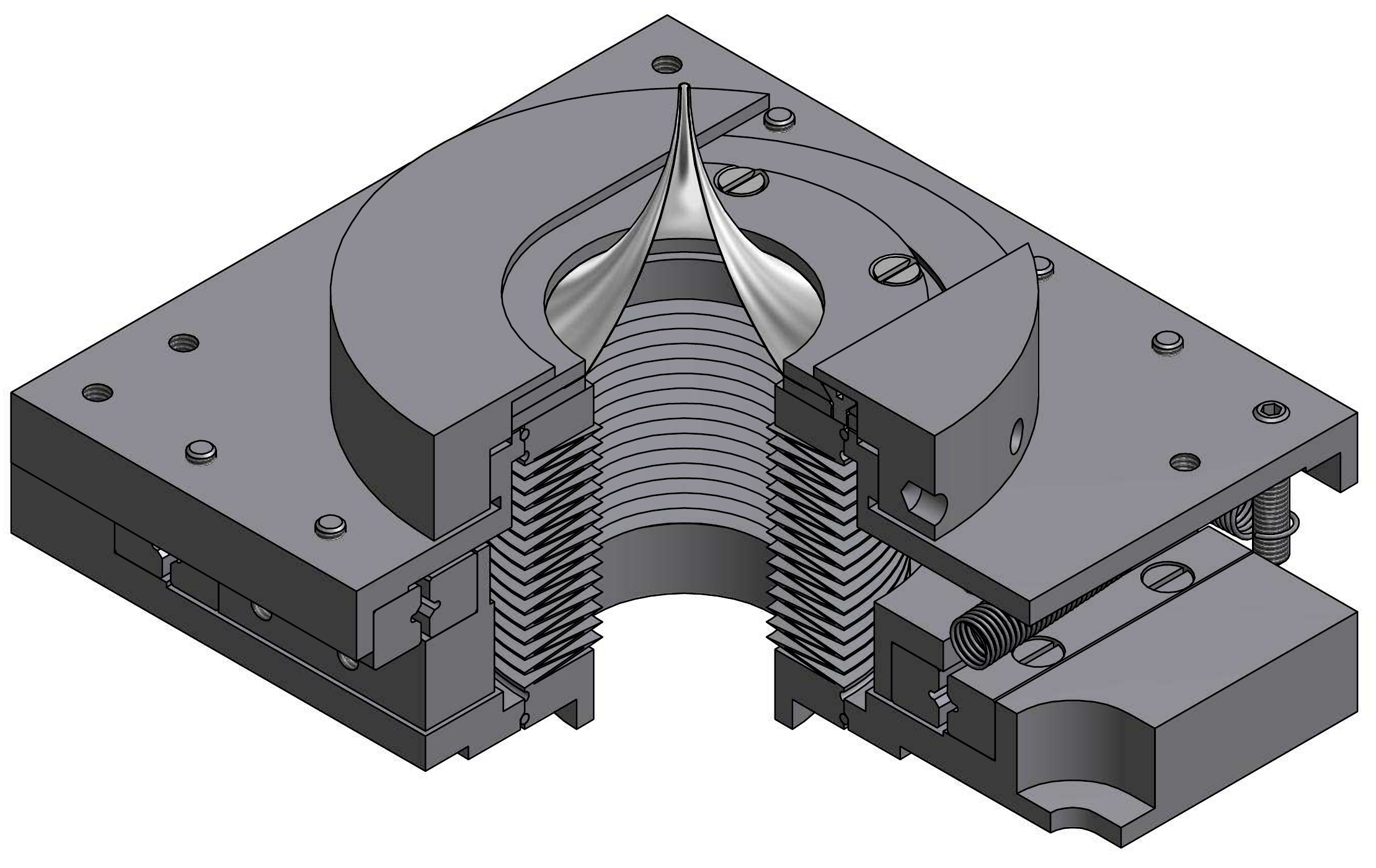}
\caption{Cross-section of the $X$-$Y$ table used to move the
skimmers. The edge-welded bellow separates the different vacuum
stages in front and behind the skimmer.} \label{fig:XYTable}
\end{figure}
\begin{figure}[t!]
\includegraphics[width=\columnwidth]{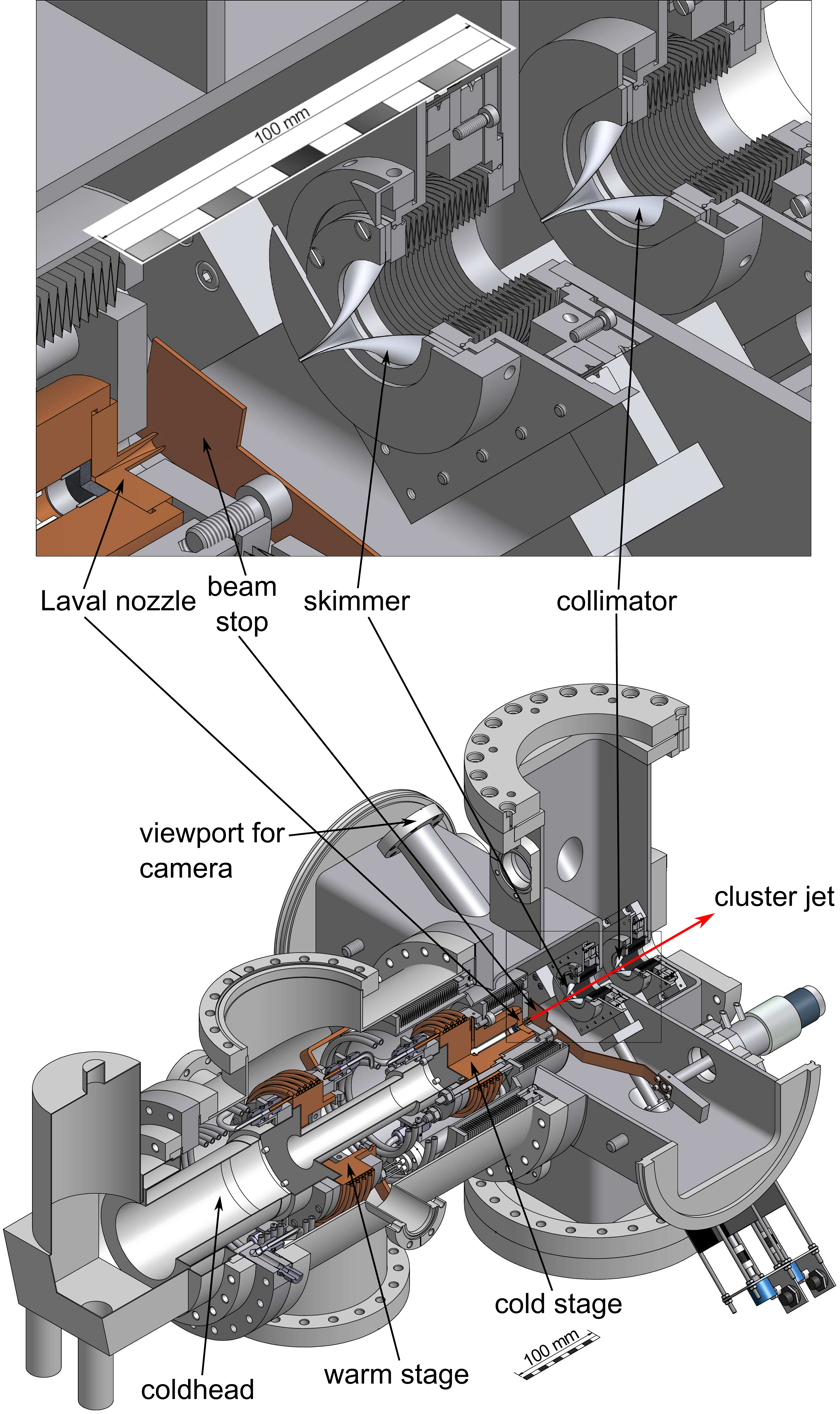}
\caption{View of the complete new cluster source and an expanded view
of the main parts of the source.} \label{fig:ClusterSource}
\end{figure}

The setup is optimized for PANDA at HESR where mostly hydrogen 
will be used as target material. The
clusters are produced by the expansion of cooled hydrogen gas or
liquid through a Laval nozzle into a vacuum chamber. The one used,
which is illustrated in Fig.~\ref{fig:Nozzle}, was manufactured in the
CERN workshop
from copper and has a minimum diameter of approximately
$28\,\mu\mathrm{m}$. The hydrogen gas, with pressures of up to
$20\,\mathrm{bar}$, has to be purified in order to prevent clogging
of the cold nozzle by frozen impurities. For this purpose a
commercial purification system (Johnson-Matthey HE20) based on a
palladium membrane is used. This achieves a purity level of
9.0, which allows for continuous operation over weeks
without any blocking of the nozzle. In addition a micro filter
with a pore size of $0.5\,\mu\mathrm{m}$ 
in the gas input line stops solid particles from entering the gas
system. The gas is cooled by a dual stage Gifford-McMahon-type cold
head (Leybold COOLPOWER 10 MD). The first stage has a cooling power
of $110\,\mathrm{W}$ at $80\,\mathrm{K}$ and the second stage
$18\,\mathrm{W}$ at $20\,\mathrm{K}$. With this
device, and a $50\,\mathrm{W}$ heating cartridge, the temperature of
the fluid directly before the nozzle can be adjusted by using a
temperature controller (LakeShore Model 331S) in the operational
region between $19$ and $50\,\mathrm{K}$. In order to minimize the
heat transfer by heat conductance from the walls of the vacuum
chamber (room temperature) to the two stages of the cold head, the
new design of the cluster source embeds the cold head in a separate
vacuum chamber (insulation vacuum chamber) which has no direct
connection to the vacuum in the other chambers of the cluster source.

The cold hydrogen passes the Laval nozzle, where the clusters are
produced during the expansion of the fluid into an adjacent vacuum
chamber, the skimmer chamber. Depending on the operational
parameters, only a small fraction of the gas produces clusters. From
this cluster jet a certain part, which depends on the experimental
requirements, is selected by skimmers. Thus most of the gas flow
through the nozzle, i.e.\ approximately up to $80\,\mathrm{mbar\cdot l/s}$ hydrogen
gas, has to be pumped away from this chamber. Since it was observed
that vacuum pressures above $\approx10^{-1}\,\mathrm{mbar}$ lead to a
loss of target thickness~\cite{Khoukaz1999}, the pumping speed of the
pumps used is chosen in order to match this requirement. Therefore,
the two pumping systems shown in Fig.~\ref{fig:VacuumSystem}, each
consisting of two roots and one rotary vane pump with a total pumping
speed of $5000\,\mathrm{m^3/h}$, are used to evacuate this vacuum
chamber.

The clusters produced can be separated from the surrounding gas by
using a trumpet-shaped skimmer with an opening diameter of the
orifice of $0.5\,\mathrm{mm}$ (Beam Dynamics Incorporation). A second
skimmer, the collimator, is mounted in an adjacent vacuum chamber.
The collimator further reduces the residual gas flow into the
accelerator and defines the shape and size of the cluster jet beam at
the interaction point. For the data presented here, a collimator with
a round orifice with an opening of $0.7\,\mathrm{mm}$ was used. Due
to the limited gas conductance of the skimmer, the vacuum chamber
where the collimator is located can be pumped by two turbomolecular
pumps. Both the skimmer and the collimator are mounted on 
step-motor-driven $X$-$Y$ tables displayed in Fig.~\ref{fig:XYTable}.
In this way it is possible to adjust them, e.g.\ for alignment
purposes during operation, in both directions perpendicular to the
cluster beam in steps of one micrometre with a maximum displacement
of two millimetres with respect to the central axis of the setup. The
complete setup is shown in Fig.~\ref{fig:ClusterSource}.
\begin{figure}
\includegraphics[width=\columnwidth]{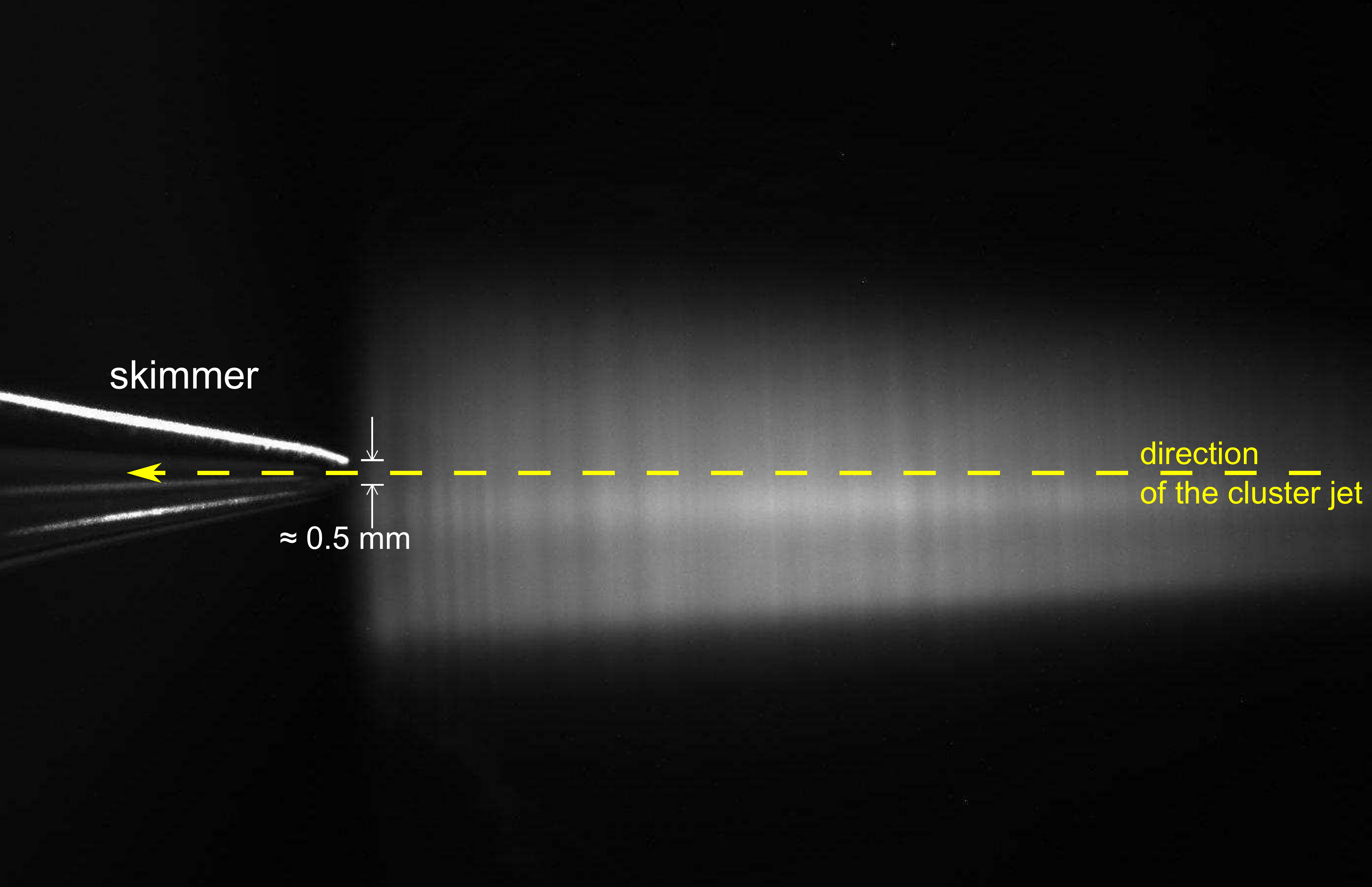}
\caption{Photograph of the cluster jet taken in the skimmer chamber.
The clusters are produced in the nozzle which is situated to the
right, outside of the range of the camera. Visible is the skimmer tip
on the left together with the cluster jet, illuminated by a laser
beam. At this stage the opening angle of the cluster jet beam is
given by the nozzle geometry.}
\label{fig:ClusterJetPhoto}
\end{figure}

A motor-driven shutter (Fig.~\ref{fig:ClusterSource}, ``beam stop'')
is installed between the nozzle exit and the orifice of the first
skimmer. This device allows one to stop the cluster jet beam
completely in the skimmer stage on a sub-second time scale. Such a
feature is of great importance for storage ring experiments where,
for each cycle, the injection, acceleration and cooling of the ion
beam are best done in the absence of a target stream.

The region between the nozzle and the skimmer can be observed by a
camera mounted outside the vacuum at an UHV vacuum window of the
skimmer chamber (Fig.~\ref{fig:ClusterSource}). If the temperature
and pressure of the incoming gas in front of the nozzle is chosen
appropriately in a region where the hydrogen is liquid, the density
can be as high that the jet is visible. An example of an image taken by this
camera is shown in Fig.~\ref{fig:ClusterJetPhoto}. The cluster jet,
which is clearly visible in this photograph, is illuminated by an
optically broadened laser beam ($P\le 1\,\mathrm{mW}$). The
vertical structures in the cluster beam are an artifact caused by the
optical system. However, the clear horizontal structures in
brightness of the jet correspond to density variations. It is
important to note that the bright core inside the beam, which is a
region of higher density, is not distributed symmetrically around the
central axis of the nozzle but is directed towards the lower part of
the picture. It was observed that the deflection angle of this core
depends strongly on the temperature and the pressure of the fluid in
front of the nozzle. At high temperatures and low pressures (e.g.\
$p=7\,\mathrm{bar}$ and $T=35\,\mathrm{K}$) the structures are almost
uniform and symmetrically arranged whereas at highest jet densities,
corresponding to low temperatures and high pressures (e.g.
$p=17\,\mathrm{bar}$ and $T=25\,\mathrm{K}$), both the deflection and
the density structure are clearly visible. This effect is currently
under investigation and indicates potential for further improvements
of the cluster beam density in the scattering
chamber~\cite{eu_proposal,protvino}.

Between the collimator chamber and the beam pipe leading to the
interaction point, two additional pumping stages, shown in
Fig.~\ref{fig:VacuumSystem}, are installed in order to minimize
further the gas flow from the source into the scattering chamber. The
first chamber is equipped with a M\"unster-type
cryopump~\cite{Dombrowski1997a} and the next one with a single
turbomolecular pump. At the exit of both chambers, a skimmer is
inserted whose inner diameter is a few millimetres larger than the
cluster beam at this position to avoid interference with the cluster
beam.

\subsection{Beam line and cluster beam dump}
\begin{figure}
\includegraphics[width=\columnwidth]{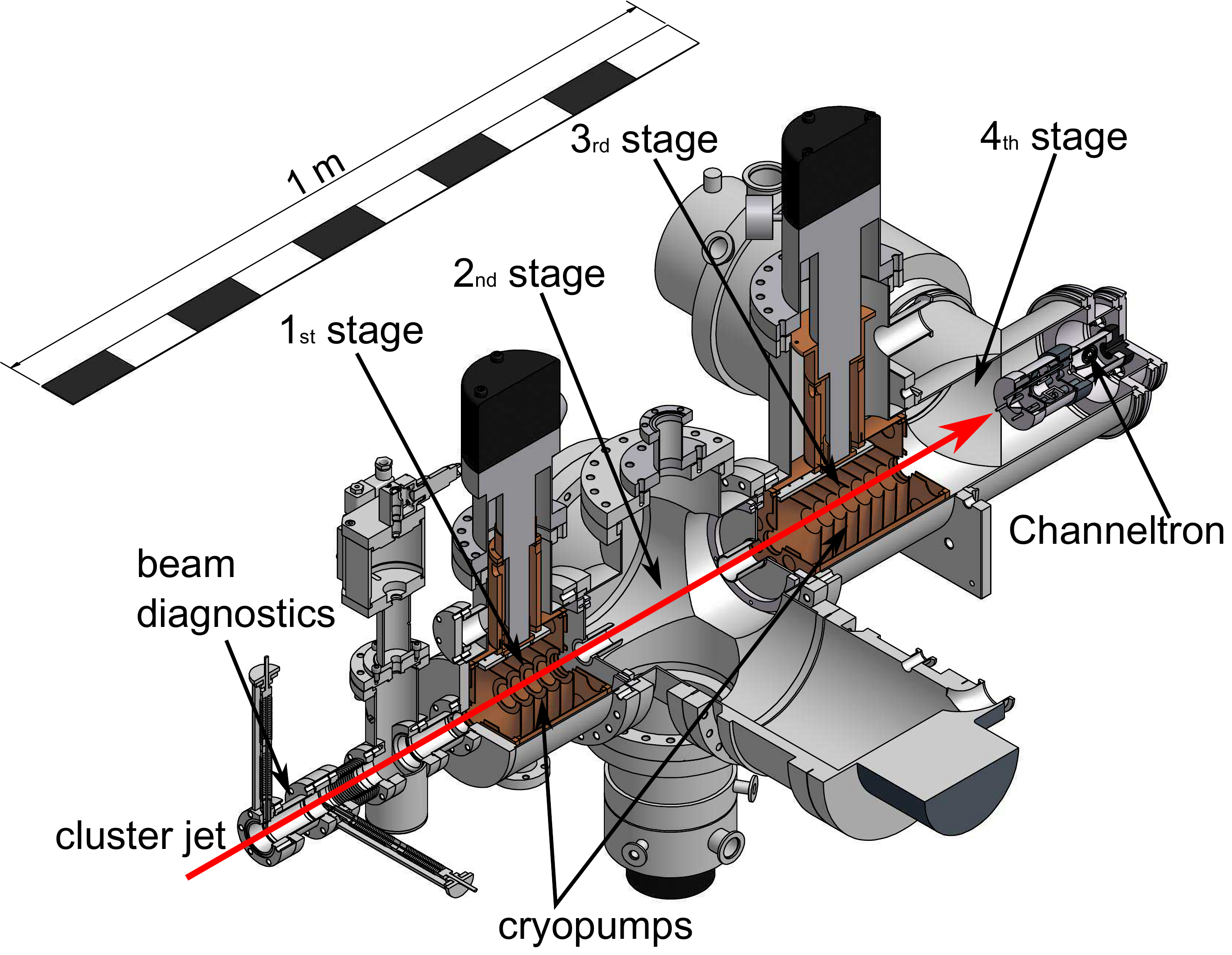}
\caption{Beam dump of the target station based on the COSY-11
design~\cite{Dombrowski1997a}.} \label{fig:ClusterBeamDump}
\end{figure}

The produced clusters are transferred through a beam pipe with an
inner diameter of $38\,\mathrm{mm}$ to the interaction region where
the scattering chamber is placed. In the scattering chamber, the
target thickness can be measured at a distance of $2.1\,\mathrm{m}$
from the nozzle using a beam diagnostic system discussed in the next
section. From this scattering chamber the clusters are transferred to
the cluster beam dump. The connecting beam pipe, with an inner
diameter of $66\,\mathrm{mm}$, is wider than that between the cluster
source and the scattering chamber. The main design of the beam dump
shown in Fig.~\ref{fig:ClusterBeamDump} matches that from the former
COSY-11 experiment~\cite{Dombrowski1997a} located at the COSY
accelerator of the Forschungszentrum J\"ulich. This beam dump
consists of four differentially pumped vacuum stages equipped with
pumps of different types. The first and third stages contain
cryopumps of the M\"unster type~\cite{Dombrowski1997a}, whereas the
other two stages are equipped with turbomolecular pumps. The complete
system is designed to minimize the back flow of the gas from the
stopped clusters into the scattering chamber. This minimization of
gas load to the scattering chamber is of special importance if the
equipment is used as an internal target at a storage ring. At the
final stage a detector system can be installed in order to measure
the cluster velocity. This system is described in detail in the
following section.

\subsection{Beam diagnostics}

\begin{figure}
\includegraphics[width=\columnwidth]{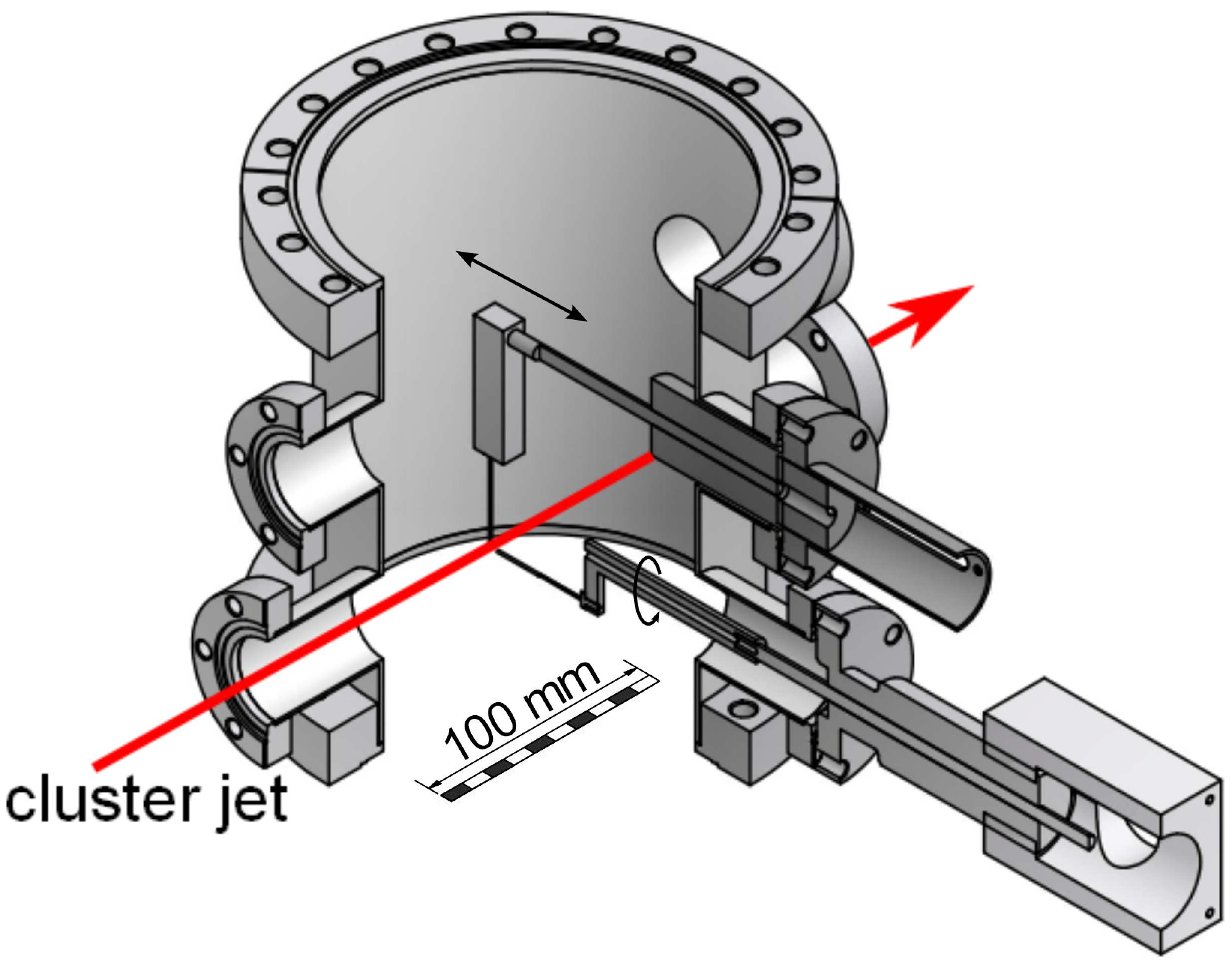}
\caption{Scattering chamber with movable rods used for the
measurement of both the absolute target thickness and the target beam
profiles.} \label{fig:ScatteringChamber}
\end{figure}

One of the most important parameters which characterizes the
performance of an internal target is the target thickness, which can
be expressed either as volume density $\rho$ or as areal density
$n_T$. In order to measure the target thickness at the interaction
point ("PANDA geometry") a vacuum chamber shown in
Fig.~\ref{fig:ScatteringChamber} was installed. In this scattering
chamber a beam diagnostic system, consisting of two movable rods, can
be used to determine the target thickness. The used rods have a
diameter of one millimetre which is much smaller than the
ten millimetre diameter of the cluster jet.
\begin{figure}
\includegraphics[width=\columnwidth]{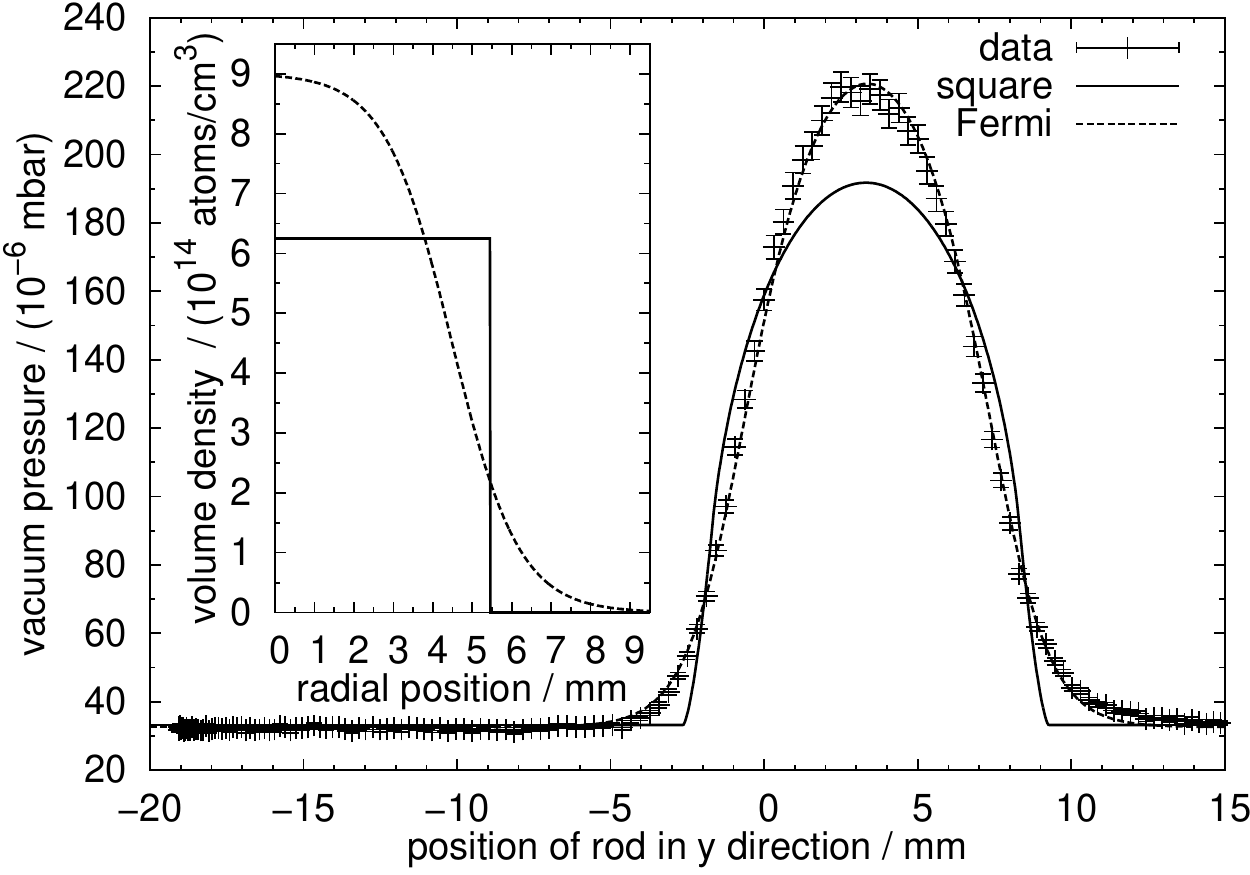}
\caption{Measured pressure in the scattering chamber as a function of
the position of the vertical rod (diameter: one millimetre) shown in
Fig.~\ref{fig:ScatteringChamber}. The solid line represents a fit
assuming a homogeneous radial volume density of the scanned cluster
jet with sharp boundaries while the dashed curve represents a fit to
a Fermi-like density distribution expressed in
Eq.~(\ref{eq:FermiDensityDistribution}). The insert shows the radial
volume density distributions corresponding to the drawn lines in the
pressure distribution.} \label{fig:PressureProfile}
\end{figure}
\begin{figure}[b!]
\includegraphics[width=\columnwidth]{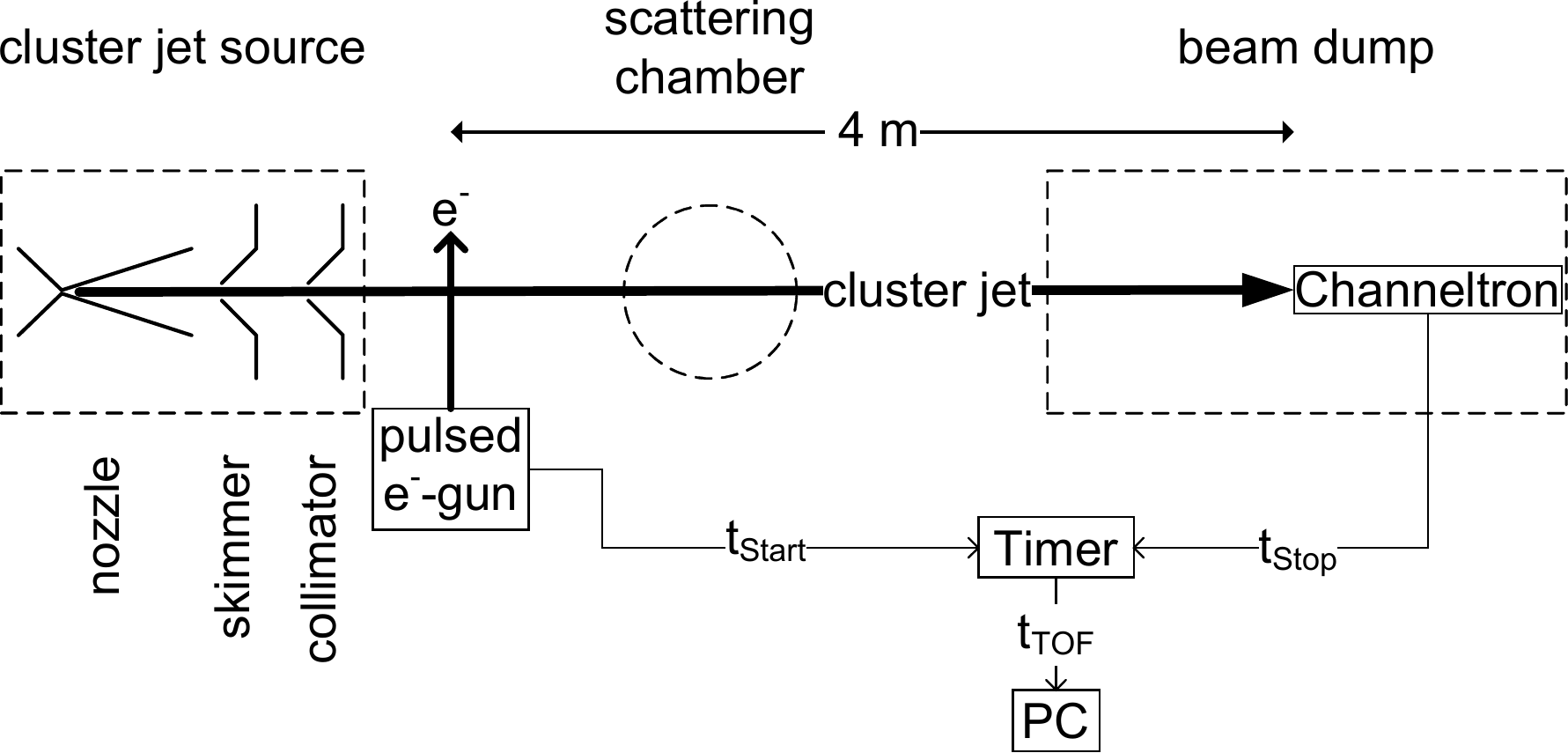}
\caption{Schematic diagram of the time-of-flight system used
to measure the velocity of individual clusters.}
\label{fig:TOFSchema}
\end{figure}
\begin{figure}[t]
\includegraphics[width=\columnwidth]{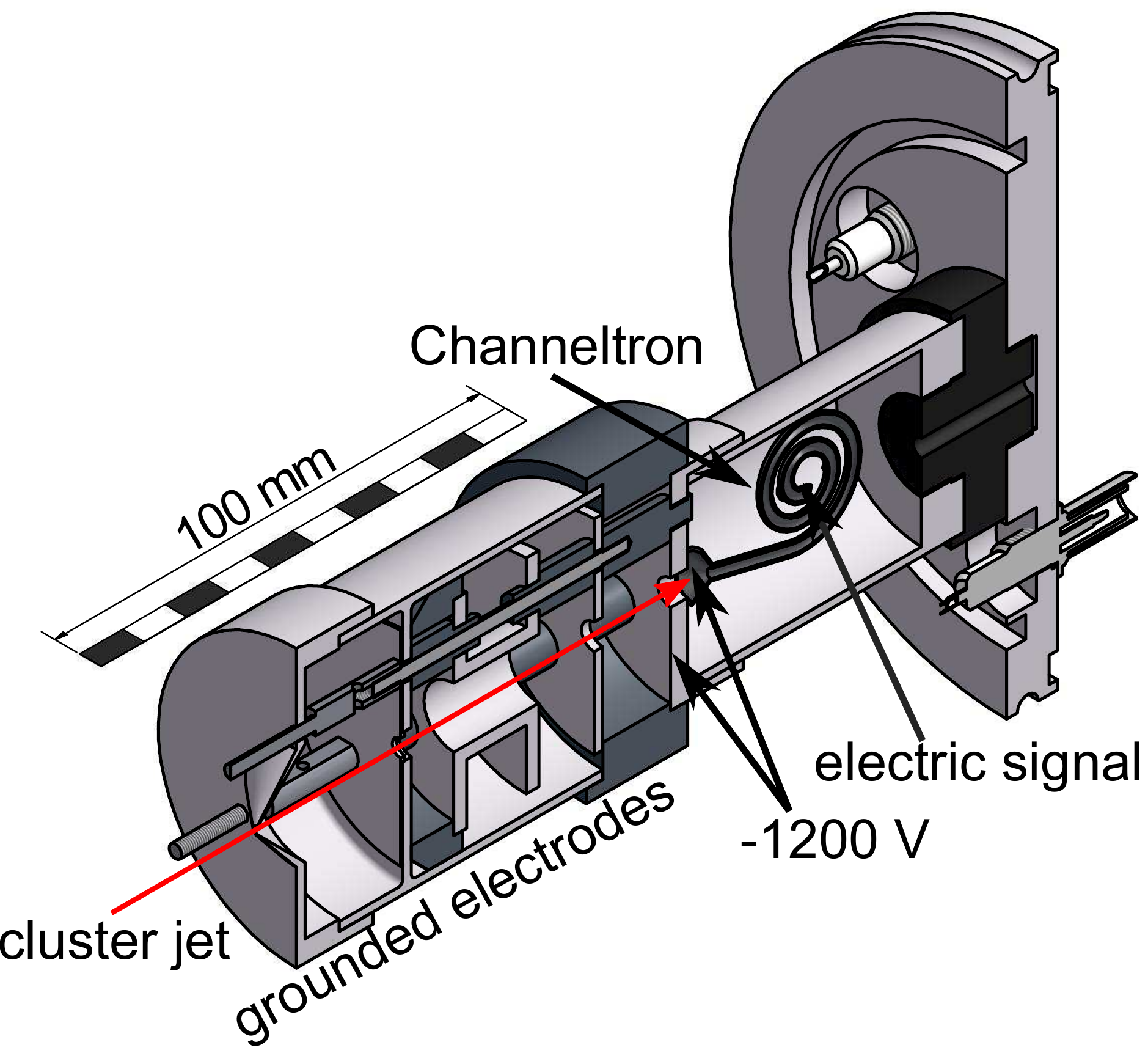}
\caption{Detection system used for the generation of the stop signal
of the TOF measurements.}
\label{fig:Channeltron}
\end{figure}
Clusters colliding with these rods break up and generate a gas load,
which increases the pressure in the vacuum chamber.
In a simple model, assuming a constant density in the
cross-section between rod and cluster jet, this pressure
increase is directly proportional to the total hydrogen
mass flow $dm/dt$ into this vacuum chamber. This allows
a direct target density determination according to:
  \begin{equation}
    \rho=\frac{dm/dt}{A\,v}
    \label{eq:massflow}
  \end{equation}
Here $A$ corresponds to the cross-section of the cluster beam in the
scattering chamber, which can be determined by beam-profile
measurements using the movable rods, and $v$ is the mean velocity of
the cluster jet. An example of such a beam profile measurement is
shown in Fig.~\ref{fig:PressureProfile}. 
Since this simple model cannot be used for the
analysis of this measured pressure increase,
the volume and areal density can be deduced by assuming a
certain volume density distribution $\rho_z(x',y')$ at the specific
distance $z$ from the nozzle. The vacuum pressure $p(x)$ as a
function of the position $x$ of the rod can be calculated from:
  \begin{equation}
    p(x) = p_\mathrm{b} + \frac{v\,R\, T}{S\, M}\int_{x-x_0-d/2}^{x-x_0+d/2}\mathrm{d}x'
           \int_{-\infty}^{+\infty}\mathrm{d}y'\,\rho_z(x',y')
    \label{eq:VacuumPressure}
  \end{equation}
where $v$ is the mean velocity of the clusters, $R$ the universal gas
constant, $T$ the temperature of the gas, $M$ the molar mass of the
hydrogen molecules, $S$ the pumping speed of the vacuum pumps used to
evacuate the scattering chamber, $d$ the diameter of the scanning
rod (here: $d=1\,\mathrm{mm}$), $x_0$ the displacement of the maximum of the volume distribution
with respect to the coordinate system of the rod, and $p_\mathrm{b}$
the background pressure in the chamber. By fitting
Eq.~(\ref{eq:VacuumPressure}) to the measured data, the volume
density can be estimated.

\begin{figure}[b!]
\includegraphics[width=\columnwidth]{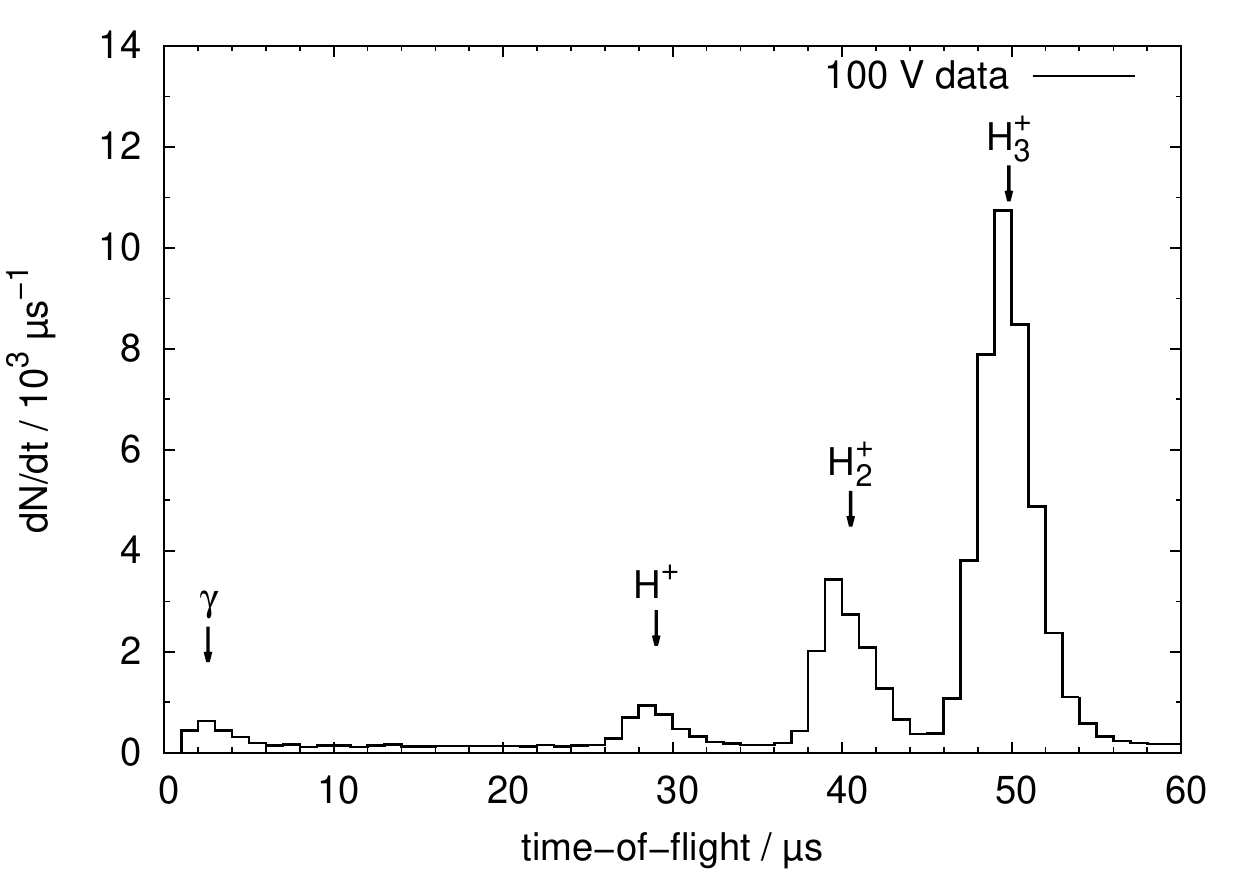}
\caption{Time-of-flight distribution of different hydrogen ions
accelerated through a voltage of $100\,\textrm{V}$. The calibration
source used to generate and accelerate the ions also emits photons
that are visible in the spectrum.}
\label{fig:CalibrationDistribution}
\end{figure}
In order to be able to use Eq.~(\ref{eq:VacuumPressure}), the mean
velocity of the clusters has to be known. For this purpose a
time-of-flight system (TOF) shown in Fig.~\ref{fig:TOFSchema} was
implemented. Behind the cluster source a pulsed electron
gun~\cite{Otte2007} is mounted, whose intensity can be reduced
sufficiently to guarantee single cluster ionization. In coincidence
with the pulse of the electron gun, a micro-controller-based timer is
started. Due to the large mass of the clusters the effect of the
ionization process on the cluster velocity can be neglected and the
ionized particles move with their original velocity to the stop
detector located in the beam dump. Here the ionized clusters are
detected by a Channeltron shown in Fig.~\ref{fig:Channeltron} which
provides the electronic signal used to stop the timer. Since neutral
clusters were found to generate no signal, spectra with minimal
background were obtained.

Before starting systematic velocity measurements, the system was
calibrated by means of TOF measurements using ionized hydrogen gas
with precisely known kinetic energy and velocity of the
ions~\cite{TaeschnerDr}. In Fig.~\ref{fig:CalibrationDistribution}
the result of such a measurement is shown. In this spectrum four
peaks can be identified, corresponding to three different hydrogen
ions, namely H$^+$, H$_2^+$, and H$_3^+$, and photons emitted from
the ionized gas. By using several measurements with different
acceleration voltages, the time offset of the electronic system used
and the length of the flight path could be determined. A length of
$4.02(3)\,\mathrm{m}$ was obtained, which agrees well with 
direct distance measurements. In addition a time resolution
of around three microseconds was observed. This originates from the
pulse width of the electron gun, as well as from the finite width of
the energy distribution of the generated ions. The time resolution is
therefore better than $0.1\%$ of the time-of-flight of a cluster, 
which is typically several milliseconds. By
using this system, the velocity distribution could be measured and
the mean velocity needed for the target thickness estimation
extracted.

\begin{figure}
\includegraphics[width=\columnwidth]{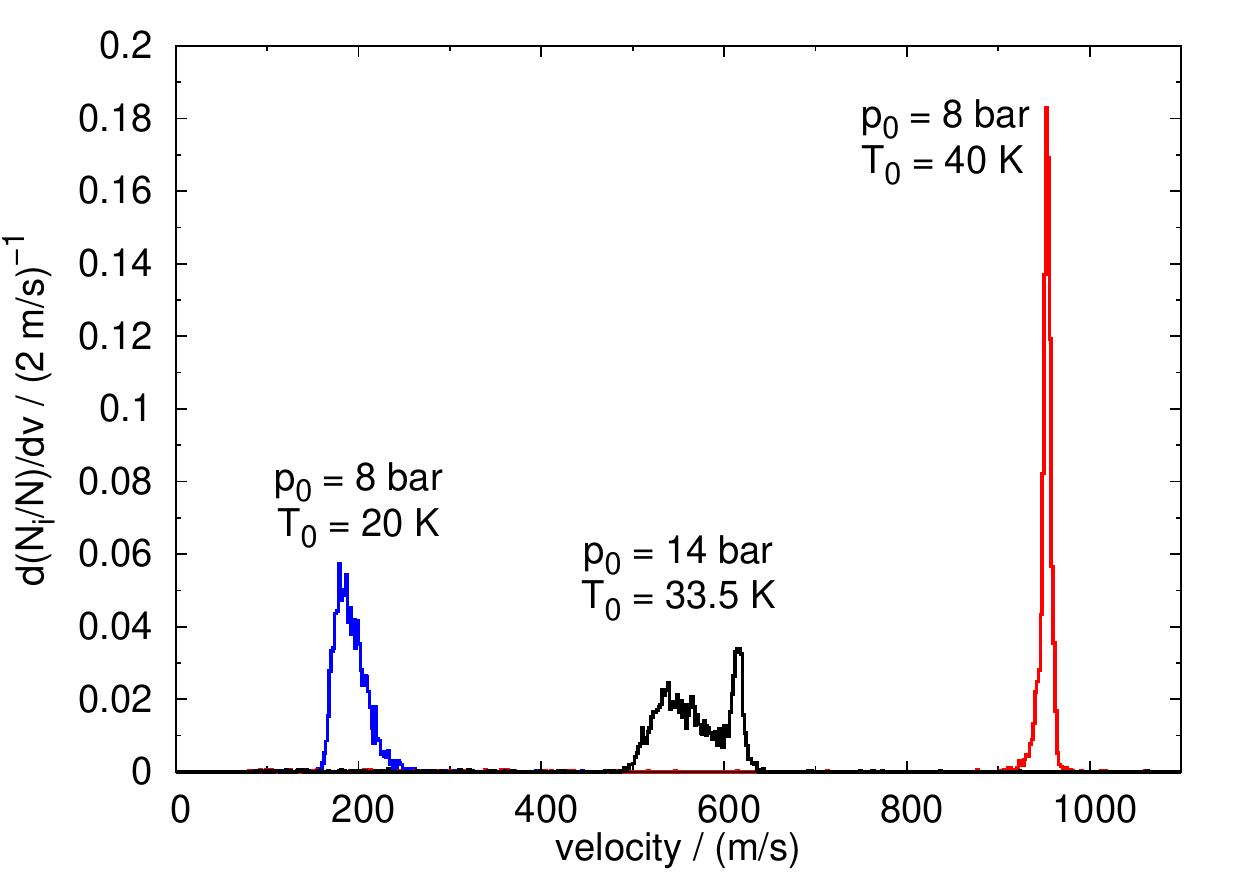}
\caption{Three examples of velocity distributions of hydrogen
clusters measured with the time-of-flight setup at different
pressures $p_0$ and temperatures $T_0$ of the hydrogen before
entering the nozzle.} \label{fig:VelocityDistributions}
\end{figure}

Examples of the measured cluster velocity spectra are shown in
Fig.~\ref{fig:VelocityDistributions}. The frequency density
distribution (ordinate) shown is obtained for three examples of
hydrogen temperature and pressure by dividing the number of
observed clusters $N_i$ in the depicted velocity intervals by the
total number of observed clusters $N=\sum{N_i}$ so that the integral
over the displayed velocity range is equal to one. The three
distributions represent different modes of operation of the
cluster jet source. At the temperature and pressure of the displayed 
distribution with the highest mean velocity of $953\,\mathrm{m/s}$,
the hydrogen is gaseous in front of the nozzle. In this case the
clusters are produced by condensation of the gas, leading to a narrow
distribution with a width ($1\sigma$) of $5\,\mathrm{m/s}$. The
distribution with the lowest mean velocity of $191\,\mathrm{m/s}$ is
observed for temperatures and pressures where the hydrogen is already
in a liquid phase before entering the nozzle. Obviously here the
clusters are produced by break up of the liquid jet injected into
vacuum. As can be seen from the spectra obtained, this production
mechanism leads to a much wider velocity distribution, on the order
of $16\,\mathrm{m/s}$ ($1\sigma$). For temperatures and pressures
near the phase transition (gas phase/liquid phase) preceding the
nozzle, a double peak structure is observed with a narrow
distribution with higher mean velocity on top of a broader
distribution. This structure was already observed in Ref.~\cite{Knuth1995}
at higher hydrogen pressures between $20$ and $120\,\mathrm{bar}$ and
interpreted as a production of clusters from a liquid/gas mixture. A
more detailed presentation of the observed distributions can be found
in~\cite{TaeschnerDr,Koehler2010}. Furthermore, quantitative
numerical calculations of the mean velocities as function of the
pressure $p_0$ and temperature $T_0$ have been
performed~\cite{TaeschnerDr,Taeschner2010}.

\section{Results}

\begin{figure}
\includegraphics[width=\columnwidth]{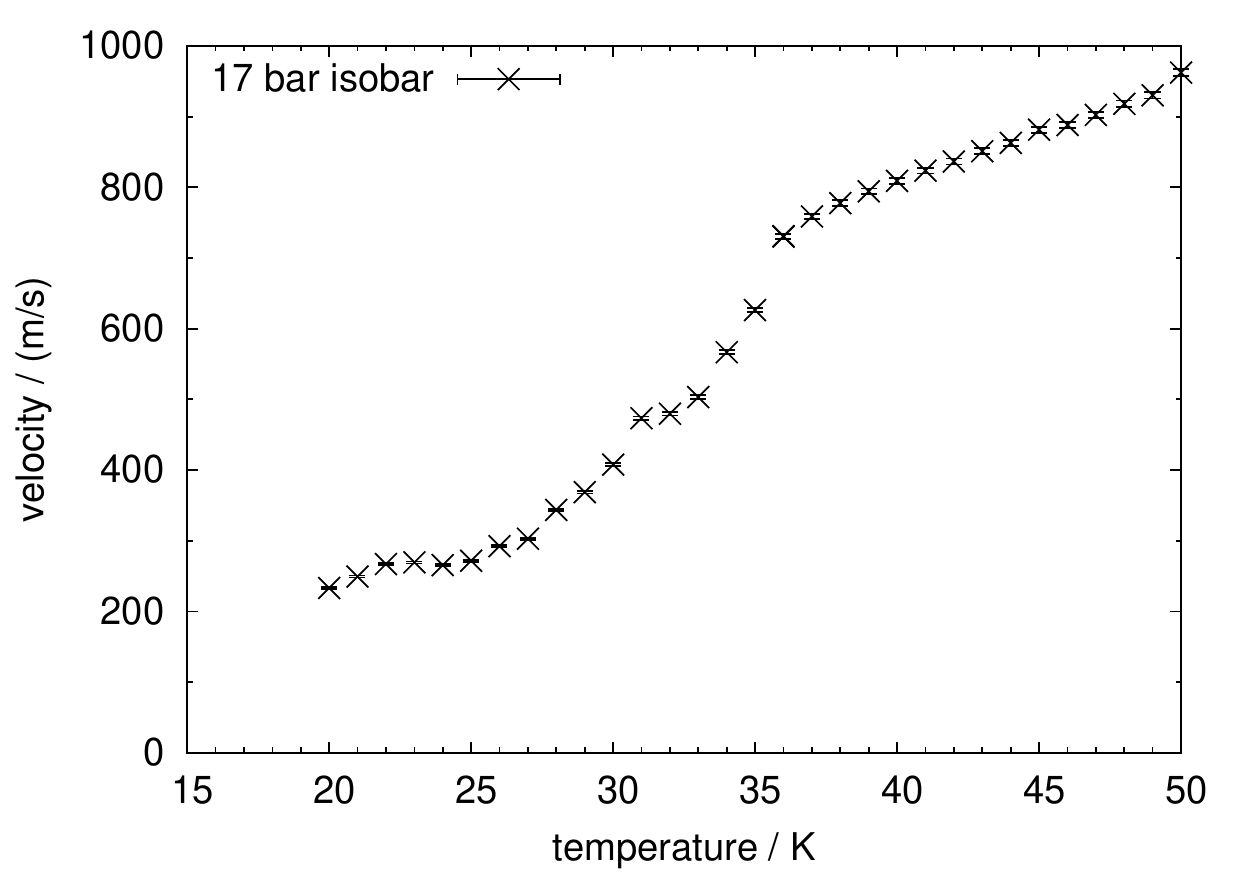}%
\caption{Mean velocity of the clusters as a function of the hydrogen
temperature at the inlet of the Laval nozzle at a constant input pressure
of $17\,\mathrm{bar}$.} \label{fig:VelocityIsobare}
\end{figure}
\begin{figure}
\includegraphics[width=\columnwidth]{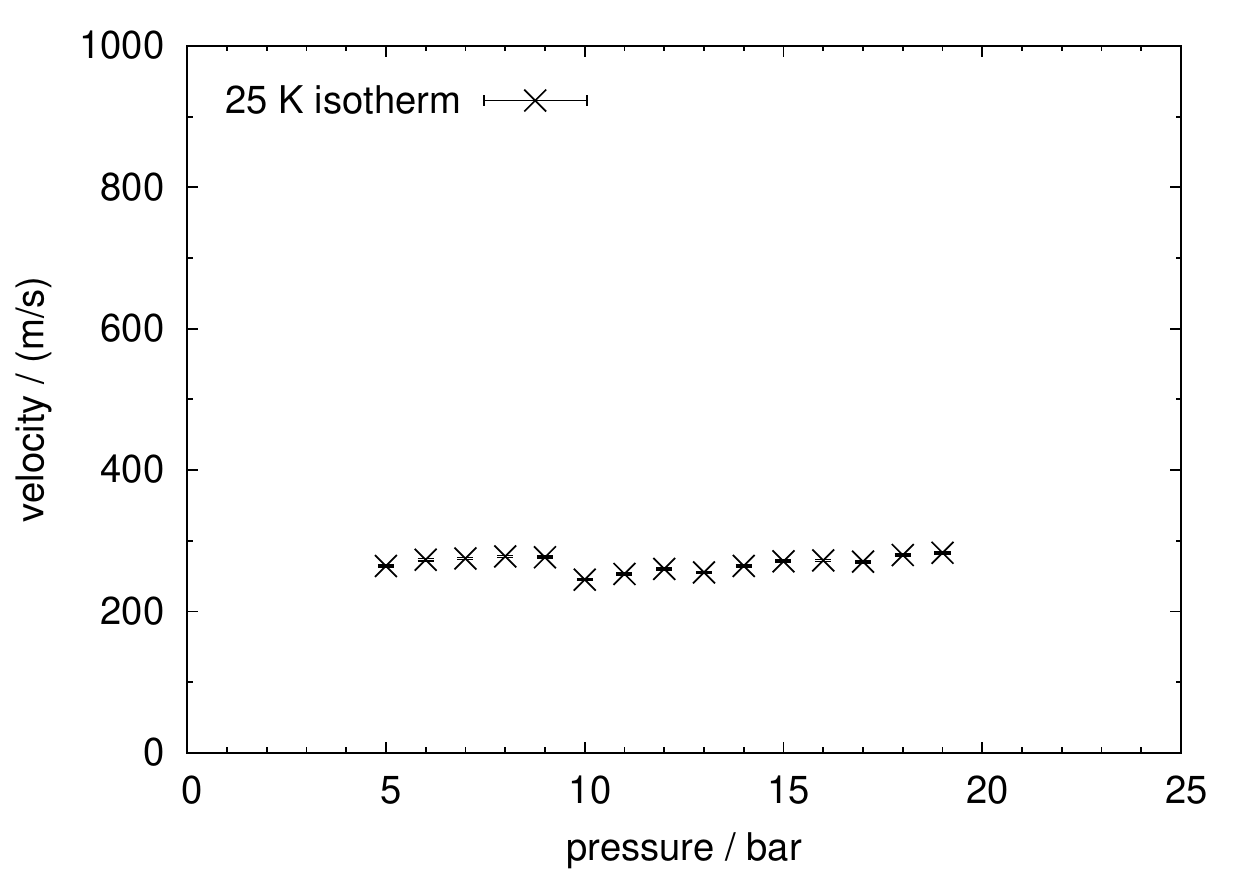}
\caption{Dependence of the mean cluster velocity on the pressure
at the inlet of the nozzle at a constant temperature of $25\,\mathrm{K}$.}
\label{fig:VelocityIsotherme}
\end{figure}
The diagnostic system described allows systematic studies on the
cluster velocities and target beam thicknesses. In
Fig.~\ref{fig:VelocityIsobare} the mean velocity is shown as a
function of the temperature of the fluid before the nozzle at a
constant pressure of $17\,\mathrm{bar}$. Since this pressure is above
the critical pressure of hydrogen,
$12.964\,\mathrm{bar}$~\cite{Leachman2009}, the phase change at the
critical temperature of $33.145\,\mathrm{K}$ is continuous. The mean
velocities displayed in Fig.~\ref{fig:VelocityIsotherme} were
measured at a constant temperature of $25\,\mathrm{K}$ as a function
of the hydrogen pressure. In both cases the uncertainties are smaller
than the symbol size.
To perform systematic measurements on cluster beam densities at the same
pressures and temperatures shown in these two graphs, the pressure
increase was recorded as a function of the position of one of the
rods in the scattering chamber. In Fig.~\ref{fig:PressureProfile} one
of these pressure profiles is presented.

In order to describe the pressure profile, two different models were
used which both assume a rotationally invariant volume density
distribution; $\rho(r)=\rho_0\,\hat{\rho}(r)$, where $\rho_0$ is the
maximum volume density and $r=\sqrt{x'^2+y'^2}$ is the distance to
the centre of the distribution, and $\hat{\rho}(r=0)=1$.
In this case the areal target thickness in units
of number of atoms per square centimetre is given by
\begin{equation}
  n_T = \frac{2\,N_A}{M_a}\,\int_0^\infty\rho(r)\,\mathrm{d}r
\end{equation}
with the molar mass of the gas atoms $M_a$ and the Avogadro constant $N_A$.
The first model represents the simplest situation and assumes a
homogeneous density distribution with a sharp edge at a radius $R$, where
\begin{equation}
  \rho_\mathrm{square}(r)= \begin{cases}
      \rho_0 & \mathrm{for}\;r\leq R \\
      0      & \mathrm{for}\;r> R.
    \end{cases}
\end{equation}
The solid line in Fig.~\ref{fig:PressureProfile} represents a fit to
the data based on this model. The corresponding volume density
distribution is shown as an inset in this figure. Obviously this
simple model is not able to describe sufficiently well the observed
density distribution. The second model uses an empiric ansatz with a
Fermi-like function to describe the volume density distribution
\begin{equation}
  \rho_\mathrm{Fermi}(r) =
  \rho_0\,\left(\exp(-\frac{R}{s})+1\right)/\left(\exp(\frac{r-R}{s})+1\right)\,,
  \label{eq:FermiDensityDistribution}
\end{equation}
with the beam radius $R$ and the parameter $s$ describing the
boundary smearing of the distribution. This type of distribution
leads to a prediction for the pressure increase shown by the dashed
line in the figure. It is clearly seen that the data can be described
with high accuracy using this kind of distribution. The reduced
values of $\chi^2/\mathrm{ndf}$ were found to be in the range between
$0.5$ and $2$.

\begin{figure}
\includegraphics[width=\columnwidth]{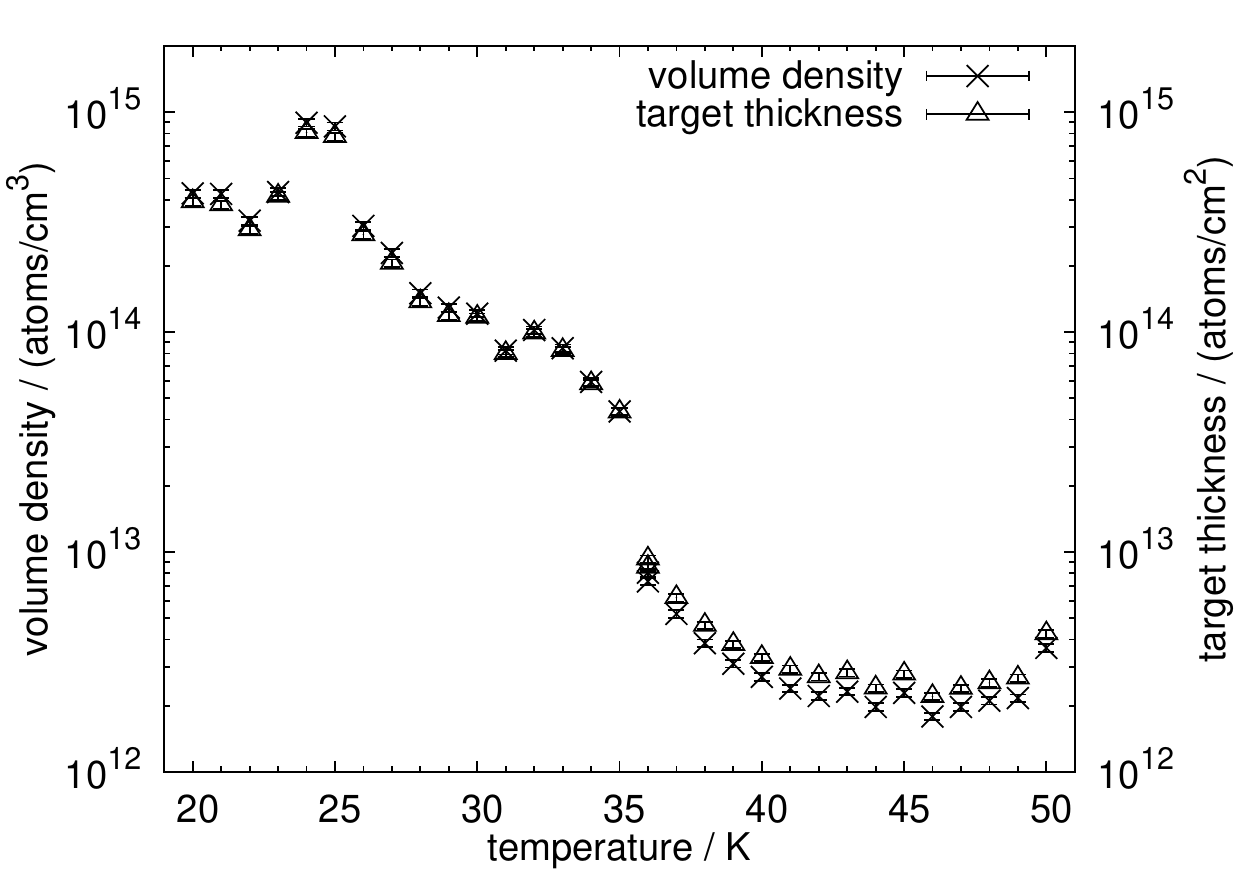}%
\caption{Volume density $\rho_0$ and target thickness $n_T$ as a function of the hydrogen temperature
at the inlet of the Laval nozzle at a constant gas pressure of
$17\,\mathrm{bar}$, assuming a Fermi-like volume
density distribution.} \label{fig:TargetThicknessIsobare}
\end{figure}
\begin{figure}
\includegraphics[width=\columnwidth]{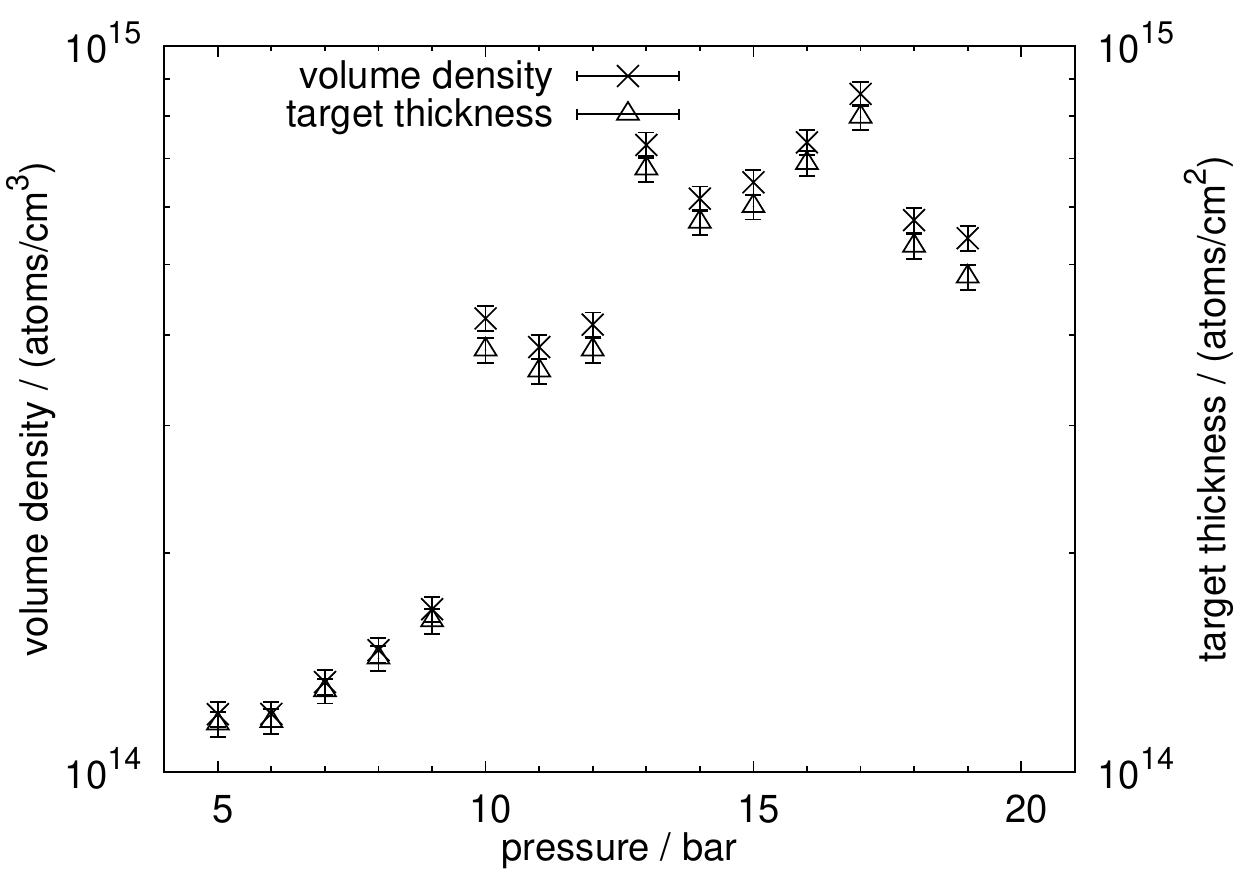}
\caption{Dependence of the volume density $\rho_0$ and target thickness $n_T$ on the pressure
at the inlet of the nozzle at a constant temperature of $25\,\mathrm{K}$, assuming a Fermi-like volume
density distribution.}
\label{fig:TargetThicknessIsotherme}
\end{figure}
Based on the parameters obtained by fitting the Fermi-like volume
density distribution to the measured data,
Figs.~\ref{fig:TargetThicknessIsobare} and
\ref{fig:TargetThicknessIsotherme} show the values found for the
maximum volume density and for the areal target thickness. At a
distance of about two metres behind the nozzle, a target thickness
of ${(8.1 \pm 0.3)\times10^{14}\,\mathrm{atoms/cm}^2}$ combined with
a target beam diameter $(2\,R)$ of approximately nine millimetres was
achieved. In agreement with earlier results, the isobaric data show a
strong correlation between the target thickness and the temperature.
The target thickness can be easily varied in a range between
$10^{12}$ and ${8\times10^{14}\,\mathrm{atoms/cm}^2}$ by choosing a
hydrogen temperature in the region between $20$ and $50\,\mathrm{K}$.
On the other hand, changes in the gas pressure do not affect the
target thickness as significantly.

Figure~\ref{fig:TargetThicknessIsotherme} shows a result of
such a measurement that was performed at a constant nozzle
temperature of $25\,\mathrm{K}$. In the pressure region between $5$
and $19\,\mathrm{bar}$, the target thickness increases by less than
one order of magnitude. These two measurements indicate the relevance
of the variation of the gas parameters e.g.\ in hadron physics
experiments. 
By adjusting the nozzle temperature one can choose the order
of magnitude of the density. Depending on the heating power and the
mass of the nozzle holder the variation of the nozzle temperature 
and the approach to a stable situation is on the order of minutes.
Instead, the density of the target can be varied within
seconds by changing the pressure of the incoming hydrogen. Therefore, 
the variation of this parameter is highly suited for a real time 
adjustment of the target thickness.  
This might
be of interest for adjustments of the luminosity in real time and can
be used to compensate for the loss of beam particles in order to
achieve a constant luminosity. The fluctuations seen in
both figures can be attributed to a change in the emission angle of
the cluster beam from the nozzle. Since these
fluctuations are reproducible and stable in time it is possible to measure
the target thickness as function of the gas pressure and the temperature at the
inlet of the nozzle. This mapping can be used for the adjustment of
the luminosity. Furthermore, a nozzle tilting system, as it is
discussed below, will allow to smooth the observed fluctuations.

Below a nozzle temperature of about
$25\,\mathrm{K}$ a brightness structure was observed in the beam, as
can be seen in Fig.~\ref{fig:ClusterJetPhoto}. Surprisingly these
structures, which are regions with different cluster jet densities,
are not symmetrically distributed with respect to the cluster beam
axis. Therefore, a dedicated mechanical adjustment system is 
currently under development~\cite{eu_proposal,protvino} which allows to tilt
the nozzle axis with respect to the skimmer axis during target
operation. Recently, in first tests using a prototype of such
a tilting system very high cluster 
beam densities have been achieved, i.e. approximately 
$2\times10^{15}\,\mathrm{atoms/cm}^3$ 
($n_T \approx 1.6\times10^{15}\,\mathrm{atoms/cm}^2$)
at a distance of two metres behind the nozzle ($p_0=18.5\,\mathrm{bar}$,  
$T_0=19\,\mathrm{K}$). This result strengthens the assumption that the
decrease of the cluster beam densities at temperature below $25\,\mathrm{K}$
(see Fig.~\ref{fig:TargetThicknessIsobare}) is caused by structures 
within the cluster jet beam and can be compensated by an online
alignment. Detailed systematic investigations on this aspect are currently 
performed.   

\section{Summary}
To perform detailed measurements on cluster beam properties a
complete cluster target installation was built in M\"unster. This
provides the possibility to test cluster jet sources in a geometry
that matches the one to be used in future experiments such as
PANDA/FAIR. By means of a time-of-flight setup based on electron
impact ionization, systematic velocity studies on individual clusters
are possible. The design and performance of a newly developed
cluster jet source was presented. This new cluster source, which is a
prototype for the PANDA experiment, is the first one that can provide
routinely a hydrogen cluster target thickness of $8\times10^{14}\,\mathrm{atoms/cm}^2$ at a
distance of two metres behind the nozzle. Together with the high
purity of its target material and the absence of a distinct time
structure of the beam, this target is ideally suited for future
internal fixed-target experiments at storage rings. Furthermore, 
an even higher target thickness of $> 10^{15}\,\mathrm{atoms/cm}^2$
has been achieved by a dedicated nozzle tilting system.
\section*{Acknowledgements}

The authors would like to thank H.~Orth for the very inspiring and
helpful discussions and H.~Baumeister and W.~Hassenmeier for their
support during the design of the target device. We are grateful to
M.~Macri and J.~Ritman for providing powerful vacuum pumps. The work
provided by the teams of our mechanical and electronic workshops is
very much appreciated and we thank them for the excellent
manufacturing of the various components. Finally we would like to
thank C.~Wilkin for suggestions regarding the manuscript. The
research project was supported by BMBF (06MS253I and 06MS9149I), GSI
F\&E program (MSKHOU1012), EU/FP6 HADRONPHYSICS (506078) and EU/FP7
HADRONPHYSICS2 (227431).







\begin{thebibliography}{00}

\bibitem{Ekstroem1995}
C.~Ekstr\"om, Nucl. Instr. and Meth. A 362 (1995) 1--15

\bibitem{Ekstroem1996}
C.~Ekstr\"om, et~al., Nucl. Instr. and Meth. A 371~(3) (1996) 572--574.

\bibitem{Allspach1998}
D.~Allspach, et~al., Nucl. Instr. and Meth. A 410~(2) (1998) 195--205.

\bibitem{Dombrowski1997a}
H.~Dombrowski, et~al., Nucl. Instr. and Meth. A 386~(2-3) (1997) 228--234.

\bibitem{Lehrach2003}
A.~Lehrach, D.~Prasuhn, {Luminosity Considerations for Internal and External
  Experiments @ COSY}, Annual Report, IKP Forschungszentrum J\"ulich (2003).

\bibitem{nordhage}
\"O.~Nordhage, Ph.D. Thesis, Uppsala Universitet, Sweden, ISSN 1651-6214, ISBN
  91-554-6649-4 (2006).

\bibitem{Taeschner2007}
A.~T\"aschner, et~al., AIP Conf. Proc. 950 (2007) 85--88.

\bibitem{Knuth1995}
E.~Knuth, F.~Schunemann, J.~P. Toennies, J. Chem. Phys. 102~(15) (1995) 6258--6271.

\bibitem{General2008}
S.~General, Diploma Thesis, Westf\"alische
  Wilhelms-Universit\"at M\"unster (2008).

\bibitem{ANKE}
S.~Barsov, et~al., Nucl. Instr. and Meth. A 462 (2001) 364--381.

\bibitem{Stein}
H.~J. Stein, et~al., Phys. Rev. ST-AB 11 (2008) 052801.

\bibitem{PANDA2005}
{PANDA Collaboration}, {Technical Progress Report - Strong Interaction Studies
  with Antiprotons}, FAIR (February 2005).

\bibitem{Taeschner2011}
A.~T\"aschner, et~al., PHN-HSD-PANDA-08, GSI Scientific Report 2010,
  GSI Report 2011-1, GSI (2010).

\bibitem{Khoukaz1999}
A.~Khoukaz, et~al., Eur. Phys. J. D 5~(2) (1999) 275--281.

\bibitem{eu_proposal}
A.~Khoukaz, Futurejet - Cryogenic Jets of Nano- and Micrometer-Sized Particles
  for Hadron Physics, HadronPhysics3: Kick-off Meeting CNRS Paris,
  17th-18th September 2010, 2010.

\bibitem{protvino}
A.~Khoukaz, Updates on Targets, PANDA Collaboration Meeting, 6th--10th June
  2011, Protvino, Russia, 2011.

\bibitem{Otte2007}
J.~Otte, Diploma Thesis, Westf\"alische
  Wilhelms-Universit\"at M\"unster (2007).

\bibitem{TaeschnerDr}
A.~T\"aschner, Doctoral Thesis, Westf\"alische
  Wilhelms-Universit\"at M\"unster, Germany (in preparation).

\bibitem{Koehler2010}
E.~K\"ohler, Diploma Thesis, Westf\"alische
  Wilhelms-Universit\"at M\"unster (2010).

\bibitem{Taeschner2010}
A.~T\"aschner, et~al., HK 7.2, DPG Fr\"uhjahrstagung, Bonn, Germany,
  March 15--19, 2010.

\bibitem{Leachman2009}
J.~W. Leachman, et~al., J. Phys. Chem. Ref. Data 38~(3) (2009) 721--748.

\bibitem{DickKubischta1988}
L.~Dick, W.~Kubischta, CERN Report (1988), CERN-EP-88-135.
				
\bibitem{Taeschner2006}
A.~T\"aschner, et~al., Annual Report, IKP Forschungszentrum J\"ulich (2006) 48.

\end{thebibliography}







\end{document}